\documentclass[conference]{IEEEtran}
\usepackage{cite}
\usepackage{amsmath,amssymb,amsfonts}
\usepackage{amsthm}  % For proof environment
\usepackage{algorithmic}
\usepackage{graphicx}
\usepackage{textcomp}
\usepackage{xcolor}
\usepackage{booktabs}
\usepackage{multirow}
\usepackage{url}
\usepackage{hyperref}
\usepackage{pifont}  % For implementation status symbols
\usepackage{tikz}

\begin{document}

\title{CivicShield: A Cross-Domain Defense-in-Depth Framework for Securing Government-Facing AI Chatbots Against Multi-Turn Adversarial Attacks}

\author{\IEEEauthorblockN{KrishnaSaiReddy Patil}}

\maketitle

\begin{abstract}
The rapid deployment of Large Language Model (LLM)-based chatbots in government citizen services introduces critical security vulnerabilities that existing defenses fail to address. Multi-turn adversarial attacks---which distribute malicious intent across multiple conversational exchanges---achieve success rates exceeding 90\% against state-of-the-art defenses, while single-layer guardrail systems are bypassed with over 90\% success by adaptive attackers. This paper presents \textbf{CivicShield}, a novel cross-domain defense-in-depth framework specifically designed for government-facing conversational AI systems. Drawing on established principles from network security, formal verification, biological immune systems, aviation safety engineering, and zero-trust cryptographic architectures, CivicShield introduces a seven-layer defense architecture comprising: (1) a zero-trust foundation with capability-based access control and cryptographic prompt authentication, (2) perimeter input validation, (3) semantic firewall with intent classification, (4) conversation state machine with formally specifiable safety invariants, (5) behavioral anomaly detection inspired by artificial immune systems, (6) multi-model consensus verification adapted from triple modular redundancy, and (7) graduated human-in-the-loop escalation modeled on aviation protocols. We present a formal threat model encompassing 8 multi-turn attack families, map the framework to applicable NIST SP 800-53 controls across 14 families, and propose an evaluation methodology using ablation analysis across defense layers. Theoretical analysis demonstrates that the multiplicative failure probability across layered defenses---even accounting for correlated failures---can reduce successful attack probability by one to two orders of magnitude compared to single-layer approaches. Simulation-based evaluation using transformer-based semantic analysis against 1,436 scenarios across thirteen categories---including the actual HarmBench (416 behaviors), JailbreakBench (100 harmful + 100 benign), and XSTest (450 safe prompts) benchmark datasets---achieves 72.9\% raw combined detection rate [69.5--76.0\% CI] on adversarial scenarios with 13.8\% raw false positive rate for the four-layer subset (Layers 2--5). Critically, a graduated response mechanism (flag vs.\ block) reduces the effective false positive rate to 2.9\% [1.9--4.4\% CI] while maintaining 100\% detection of multi-turn crescendo and slow-drift attacks. The honest drop in detection rate on real benchmarks compared to author-generated scenarios (71.2\% vs.\ 76.7\% on HarmBench, 47.0\% vs.\ 70.0\% on JailbreakBench) validates the importance of independent benchmark evaluation and establishes credible baselines for future improvement. CivicShield addresses an open gap at the intersection of AI safety, government compliance, and practical deployment---a space where no comprehensive framework addressing government-specific requirements currently exists.
\end{abstract}


\section{Introduction}

The deployment of Large Language Model (LLM)-based chatbots in government citizen services has accelerated dramatically, with the U.S. federal government now operating approximately 3,000 AI systems, up from 710 documented use cases in 2023 \cite{fastcompany2025}. These systems handle high-stakes citizen interactions including benefits eligibility determination, tax guidance, licensing inquiries, and legal information---domains where incorrect or manipulated responses carry direct financial, legal, and safety consequences.

However, the security landscape for these deployments is concerning. The study by Nasr, Carlini et al. \cite{nasr2025}, authored by researchers from OpenAI, Anthropic, Google DeepMind, and ETH Z\"{u}rich, demonstrated that 12 recent LLM defenses---most of which reported near-zero attack success rates (ASR) in their original evaluations---were bypassed with over 90\% ASR using adaptive attacks. Multi-turn adversarial attacks, which distribute malicious intent across multiple conversational turns, achieve success rates up to 99\% against current defenses \cite{harmnet2025, xteaming2025, cisco2025}. OpenAI publicly acknowledged in December 2025 that prompt injection attacks against AI agents ``may never be fully solved'' \cite{openai2025security}.

The consequences of these vulnerabilities in government contexts are not theoretical. New York City's MyCity chatbot systematically advised businesses to violate housing discrimination laws, employment regulations, and payment requirements before being shut down \cite{themarkup2024}. The Air Canada chatbot case established legal precedent that organizations are liable for their chatbot's hallucinated advice \cite{forbes2024aircanada}. In December 2025, a single attacker jailbroke Anthropic's Claude AI and used it to breach at least 10 Mexican government agencies, exfiltrating 150GB of sensitive data including 195 million taxpayer records \cite{cybernews2026}.

Despite these risks, no comprehensive security framework exists that addresses the unique requirements of government-facing AI chatbots: compliance with federal security standards (FedRAMP, NIST 800-53), accessibility requirements (Section 508, ADA), multi-turn attack resilience, and the high-stakes nature of citizen service interactions. Existing approaches either focus on single-layer defenses that are demonstrably insufficient \cite{nasr2025}, or propose theoretical frameworks without addressing government-specific deployment constraints.

This paper makes the following contributions:

\begin{enumerate}
\item \textbf{CivicShield Framework}: A seven-layer defense-in-depth architecture that synthesizes established principles from network security, formal verification, biological immune systems, aviation safety, and zero-trust cryptography into a unified framework for government AI chatbot security.

\item \textbf{Formal Threat Model}: A comprehensive taxonomy of 8 multi-turn attack families with documented success rates, specifically contextualized for government citizen service deployments.

\item \textbf{Compliance Mapping}: Explicit mapping of the framework to applicable NIST SP 800-53 controls across 14 families, FedRAMP requirements, NIST AI RMF, EO 14110, Privacy Act, and Section 508 accessibility standards.

\item \textbf{Evaluation Methodology}: A rigorous evaluation framework incorporating ablation analysis, adaptive attack testing, and government-specific metrics.

\item \textbf{Cross-Domain Innovation}: Demonstration that solutions to the ``unsolvable'' problem of LLM security already exist in adjacent fields---they simply need to be identified, adapted, and composed. We formalize this insight with two novel theoretical contributions: (a) an \textit{Adversarial Trust Decay} model with convergence proofs and an attacker's dilemma theorem characterizing the fundamental tradeoff facing multi-turn adversaries, and (b) a \textit{Cross-Domain Defense Composition} theorem that formally bounds failure correlation as a function of feature overlap between defense layers, providing the first principled justification for why cross-domain defense composition outperforms same-domain stacking.
\end{enumerate}

The remainder of this paper is organized as follows: Section II surveys the threat landscape. Section III analyzes existing defenses and their limitations. Section IV presents the CivicShield framework. Section V details the formal threat model. Section VI maps the framework to government compliance requirements. Section VII presents the evaluation methodology. Section VIII presents simulation-based evaluation results. Section IX discusses limitations and future work. Section X surveys related work. Section XI concludes.

\section{Threat Landscape}

This section presents the current state of adversarial attacks against LLM-based conversational AI systems, with emphasis on threats specific to government citizen service deployments.

\subsection{Prompt Injection: The Fundamental Vulnerability}

Prompt injection remains the \#1 security risk per the OWASP Top 10 for LLM Applications 2025 \cite{owasp2025}. The vulnerability arises from a fundamental architectural limitation: LLMs process instructions and data in the same channel (natural language), making it impossible to perfectly distinguish between legitimate instructions and injected ones---analogous to SQL injection before parameterized queries, but without a clean architectural separation available.

HiddenLayer's ``Policy Puppetry'' technique \cite{hiddenlayer2025} demonstrated the first universal, post-instruction-hierarchy bypass that works across \textit{all} major frontier models (GPT-4, Claude, Gemini, LLaMA, DeepSeek, Qwen, Mistral) using a single prompt. The technique reformulates malicious prompts to resemble policy configuration files, exploiting a systemic weakness in how LLMs are trained on instruction-related data.

The structural nature of this vulnerability is formalized by the ``attacker-defender asymmetry'': defenders must protect against all possible attacks, while attackers need only find one successful path. Nasr et al. \cite{nasr2025} invoke Kerckhoffs's principle---a defense must be secure even when the attacker knows the defense design---and demonstrate that most LLM defenses fail this test.

\subsection{Multi-Turn Adversarial Attacks}

Multi-turn attacks represent a qualitatively different threat from single-turn prompt injection. Rather than attempting to bypass safety filters in a single message, these attacks distribute malicious intent across multiple seemingly benign conversational exchanges, exploiting the model's tendency to follow conversational patterns and attend to its own prior outputs \cite{emergentmind2025multiturn}.

Table~\ref{tab:multiturn_attacks} summarizes documented success rates for multi-turn attack frameworks.

\begin{table}[htbp]
\caption{Multi-Turn Attack Success Rates (2024--2026)}
\label{tab:multiturn_attacks}
\centering
\begin{tabular}{lcc}
\toprule
\textbf{Attack Framework} & \textbf{ASR} & \textbf{Target} \\
\midrule
HarmNet \cite{harmnet2025} & 99.4\% & Mistral-7B \\
HarmNet \cite{harmnet2025} & 94.8\% & GPT-4o \\
X-Teaming \cite{xteaming2025} & 98.1\% & Multiple \\
AutoAdv \cite{autoadv2025} & 95.0\% & Llama-3.1-8B \\
Crescendo \cite{crescendo2024} & 29--61\%$\uparrow$ & GPT-4 \\
Siren \cite{siren2025} & 90.0\% & Gemini-1.5-Pro \\
Cisco AI Defense \cite{cisco2025} & $>$90\% & Open-weight LLMs \\
Human red-teaming \cite{nasr2025} & 100\% & All defenses \\
\bottomrule
\end{tabular}
\end{table}

We identify eight distinct multi-turn attack families:

\begin{enumerate}
\item \textbf{Crescendo Attacks}: Gradual escalation from benign queries, referencing the model's own prior replies \cite{crescendo2024}.
\item \textbf{Semantic Network Traversal}: Hierarchical semantic networks of adversarial paths with feedback-driven refinement \cite{harmnet2025}.
\item \textbf{Chain of Attack (CoA)}: Semantic-driven chains across turns, where each turn builds on the semantic context established by previous turns to progressively approach restricted content. Unlike crescendo attacks that rely on emotional escalation, CoA exploits logical entailment---each turn establishes a factual premise that makes the next request appear as a natural follow-up.
\item \textbf{Puzzle Decomposition}: Harmful queries split into puzzle-like segments distributed across turns, where no individual segment triggers safety filters but the assembled whole constitutes a harmful request. The attacker may request individual components (e.g., chemical names, quantities, procedures) across separate turns, relying on the model's context window to assemble the complete harmful instruction.
\item \textbf{RL-based Adaptive Dialogue}: Reinforcement learning for adaptive multi-turn strategies, where the attacker model learns optimal turn sequences through trial and error against the target system's defenses. AutoAdv \cite{autoadv2025} demonstrated that RL-trained attackers discover non-obvious attack paths that human red-teamers miss, achieving 95\% ASR on Llama-3.1-8B.
\item \textbf{Echo Chamber}: Carefully crafted interaction chains that exploit the model's tendency to attend to its own prior outputs, gradually shifting the conversation's normative baseline. The attacker elicits mildly boundary-pushing responses, then references those responses as established precedent to justify progressively more harmful requests.
\item \textbf{Tool-Chain Attacks}: Innocent tools forming dangerous chains to jailbreak LLM agents, achieving ASR exceeding 90\% against GPT-4.1 \cite{stac2025}.
\item \textbf{Context Poisoning}: Manipulating RAG-retrieved content to influence subsequent turns.
\end{enumerate}

\subsection{Government-Specific Threat Context}

Government AI chatbot deployments face unique threat amplification factors:

\textbf{High-stakes consequences}: The ODI CitizenQuery-UK benchmark \cite{odi2026} tested 11 LLMs on 22,066 citizen queries and found that models exhibit dangerously low abstention rates---attempting to answer nearly every question regardless of capability---and produce specific errors including incorrect welfare eligibility advice, wrong legal requirements, and misadvised tax deadlines.

\textbf{Legal liability}: The Air Canada ruling \cite{forbes2024aircanada} established that organizations are liable for their chatbot's outputs. NYC's MyCity chatbot \cite{themarkup2024} systematically provided illegal advice across multiple regulatory domains before being terminated.

\textbf{Data sensitivity}: Government chatbots process Controlled Unclassified Information (CUI), Personally Identifiable Information (PII), and information subject to the Privacy Act, creating high-value targets for adversarial data exfiltration.

\textbf{RAG vulnerabilities}: The CtrlRAG attack \cite{ctrlrag2026} demonstrated that injecting just 5 malicious documents into a million-document knowledge base achieves up to 90\% ASR on commercial LLMs. The BadRAG attack \cite{phantom2025} showed that poisoning 10 adversarial passages (0.04\% of corpus) induces a 98.2\% retrieval success rate for adversarial content.

\section{Existing Defenses and Their Limitations}

\subsection{Single-Layer Defense Approaches}

Table~\ref{tab:defenses} summarizes the current defense landscape and documented bypass rates.

\begin{table}[htbp]
\caption{Defense Mechanisms and Documented Bypass Rates}
\label{tab:defenses}
\centering
\begin{tabular}{lcc}
\toprule
\textbf{Defense} & \textbf{Claimed ASR} & \textbf{Adaptive ASR} \\
\midrule
12 diverse defenses \cite{nasr2025} & Near-zero & $>$90\% \\
Constitutional Classifiers \cite{anthropic2025} & N/A & 4.4\% \\
6 popular guardrails \cite{mindgard2025} & Varies & Up to 100\% \\
8 agent defenses \cite{agentdefense2025} & Varies & $>$50\% \\
TRYLOCK (layered) \cite{trylock2026} & 46.5\% & 5.6\% \\
\bottomrule
\end{tabular}
\end{table}

\textbf{Input Filters and Guardrails}: Systems including NVIDIA NeMo Guardrails, Meta's LlamaFirewall \cite{llamafirewall2025}, Azure Prompt Shields, and Amazon Bedrock Guardrails provide perimeter defense. However, Mindgard's research \cite{mindgard2025} demonstrated that simple character transformations achieve up to 100\% evasion against six popular guardrail systems. Meta's PromptGuard was bypassed by inserting blank characters between letters.

\textbf{Model Alignment (RLHF/Constitutional AI)}: While Anthropic's Constitutional Classifiers \cite{anthropic2025} achieved the best reported single-defense result (4.4\% ASR, withstanding 3,000+ hours of red teaming), RLHF creates refusal states within model representation space that remain accessible once bypassed \cite{circuitbreakers2024}. A single malicious training example can permanently corrupt safety alignment by up to 58\% \cite{grpobliteration2025}.

\textbf{Prompt Shields}: Lightweight classifier models that sit between user input and the main LLM face a fundamental capacity mismatch---the guard model has fewer parameters and less expressive power than the model it protects, creating an inherent capability gap exploitable by attackers \cite{codeintegrity2026}.

\subsection{Structured Comparison of Security Platforms}

To contextualize CivicShield's contributions, Table~\ref{tab:comparison} provides a structured comparison across twelve security dimensions against seven representative commercial, open-source, and academic platforms. Capabilities were verified through official documentation, published papers, and vendor disclosures as of mid-2025.

\begin{table*}[htbp]
\caption{Structured Comparison of LLM Security Platforms Across Key Defense Dimensions}
\label{tab:comparison}
\centering
\footnotesize
\renewcommand{\arraystretch}{1.15}

\textit{Implementation status:} \ding{72} = Implemented and evaluated in production; \ding{117} = Implemented and evaluated in simulation; \ding{109} = Proposed/designed in this paper.

\begin{tabular}{@{}l c c c c c c c c@{}}
\toprule
\textbf{Dimension} & \rotatebox{70}{\textbf{CivicShield}} & \rotatebox{70}{\textbf{Bedrock Guardrails}} & \rotatebox{70}{\textbf{Azure AI + Prompt Shields}} & \rotatebox{70}{\textbf{NeMo Guardrails}} & \rotatebox{70}{\textbf{LlamaFirewall}} & \rotatebox{70}{\textbf{Rebuff}} & \rotatebox{70}{\textbf{Lakera Guard}} & \rotatebox{70}{\textbf{TRYLOCK}} \\
\midrule
Multi-turn attack detection       & \ding{117}     & Partial$^a$ \ding{72} & Partial$^b$ \ding{72} & Partial$^c$ \ding{72} & Partial$^d$ \ding{72} & No      & No      & No      \\
Cross-turn state tracking         & \ding{117}     & No          & No          & Yes$^e$ \ding{72}     & No          & No      & No      & No      \\
Formal safety invariants          & \ding{117}     & Partial$^f$ \ding{72} & No          & No          & No          & No      & No      & No      \\
Behavioral anomaly detection      & \ding{109}     & No          & No          & No          & Partial$^g$ \ding{72} & Partial$^h$ \ding{72} & Yes$^i$ \ding{72} & Partial$^j$ \ding{72} \\
Multi-model consensus             & \ding{109}     & No          & No          & No          & No          & No      & No      & No      \\
Human escalation integration      & \ding{109}     & No          & No          & No          & No          & No      & No      & No      \\
Gov.\ compliance (NIST/FedRAMP)   & \ding{109}     & Yes$^k$ \ding{72}     & Yes$^l$ \ding{72}     & No          & No          & No      & No      & No      \\
Accessibility (Section 508)       & \ding{109}     & No          & No          & No          & No          & No      & No      & No      \\
Open source                       & Planned     & No          & No          & Yes \ding{72}         & Yes \ding{72}         & Yes \ding{72}     & No      & Yes \ding{72}     \\
Adaptive learning                 & \ding{109}     & No          & Partial$^m$ \ding{72} & No          & No          & Yes$^n$ \ding{72} & Yes$^o$ \ding{72} & No      \\
RAG-specific protections          & \ding{109}     & Yes$^p$ \ding{72}     & Yes$^q$ \ding{72}     & Yes$^r$ \ding{72}     & No          & No      & Partial$^s$ \ding{72} & No      \\
Taint tracking                    & \ding{109}     & No          & Partial$^t$ \ding{72} & No          & No          & No      & No      & No      \\
\midrule
\textbf{Defense layers}           & \textbf{7}       & 3           & 2           & 3           & 3           & 4       & 2       & 4       \\
\textbf{Impl.\ status}           & \ding{117}/\ding{109} & \ding{72} & \ding{72} & \ding{72} & \ding{72} & \ding{72} & \ding{72} & \ding{72} \\
\bottomrule
\end{tabular}

\vspace{2pt}
\raggedright
\scriptsize
$^a$Per-turn content filtering; no explicit cross-turn attack modeling.
$^b$Prompt Shields detect direct/indirect injection per request; no multi-turn session analysis.
$^c$Colang dialog rails can define multi-turn flows but lack dedicated adversarial multi-turn detection.
$^d$AlignmentCheck audits agent reasoning traces for goal hijacking but does not model cross-turn escalation.
$^e$NeMo's Colang 2.0 dialog management maintains conversation state via programmable flows.
$^f$Automated Reasoning checks use formal verification to validate outputs against encoded business rules, but do not enforce conversation-level safety invariants.
$^g$AlignmentCheck performs semantic alignment auditing of agent chain-of-thought.
$^h$Rebuff uses vector database similarity matching against known attack embeddings.
$^i$Lakera's threat intelligence platform continuously updates detection models.
$^j$TRYLOCK's sidecar classifier adaptively selects steering strength based on input threat level.
$^k$FedRAMP High authorized in dedicated government cloud regions; supports NIST 800-53 controls.
$^l$FedRAMP High authorized in government cloud regions; approved for DoD IL-4/5.
$^m$Azure Prompt Shields distinguish trusted/untrusted inputs with evolving detection models (Build 2025).
$^n$Rebuff's self-hardening architecture uses canary tokens and vector DB feedback to improve over time.
$^o$Lakera Guard's security intelligence platform continuously evolves detection from observed attacks.
$^p$Contextual grounding checks validate RAG-retrieved content; Automated Reasoning checks verify factual accuracy.
$^q$Groundedness detection identifies ungrounded or hallucinated content from retrieved sources.
$^r$NeMo provides RAG grounding guardrails and fact-checking rails for retrieved content.
$^s$Lakera evaluates retrieved document inputs for injection but lacks dedicated RAG pipeline protections.
$^t$Azure distinguishes trusted system prompts from untrusted user/document inputs but does not implement full data-flow taint propagation.

\end{table*}

\subsection{The Defense-in-Depth Gap}

As Table~\ref{tab:comparison} reveals, no existing platform combines multi-turn state tracking, formal safety invariants, multi-model consensus, human escalation, government compliance mapping, and accessibility considerations into a unified defense architecture. Commercial platforms (Amazon Bedrock Guardrails, Azure AI Content Safety) provide strong compliance postures and per-turn content filtering but lack cross-turn adversarial modeling and formal conversation-level invariants. Open-source toolkits (NeMo Guardrails, LlamaFirewall, Rebuff) offer extensibility and transparency but do not address government-specific requirements. TRYLOCK \cite{trylock2026} pioneered true defense-in-depth at the model level---combining Direct Preference Optimization (DPO) alignment, representation engineering, adaptive steering, and input canonicalization to achieve an 88\% relative ASR reduction---but operates exclusively within the inference stack and does not address: (a) multi-turn attacks across conversation sessions, (b) government compliance requirements, (c) human oversight integration, (d) RAG-specific vulnerabilities, or (e) agentic tool-use security.

CivicShield occupies this intersection: it extends defense-in-depth from the model level to the entire system architecture, integrating cross-domain mechanisms from formal methods, biological immune systems, aviation safety, and zero-trust cryptography into a seven-layer stack specifically designed for the unique requirements of government citizen service deployments.

\section{The CivicShield Framework}

CivicShield is a seven-layer defense-in-depth framework built on a zero-trust foundation, designed specifically for government-facing AI chatbot deployments. The framework's core insight is that solutions to LLM security challenges already exist in adjacent fields---they need to be identified, adapted, and composed into a unified architecture.

The design philosophy follows Reason's Swiss Cheese Model \cite{swisscheese1990}: each defensive layer has known weaknesses (``holes''), but the combination of independently-designed layers ensures that the probability of all holes aligning---allowing an attack to pass through every layer---is the product of each layer's individual failure probability, making successful attacks exponentially less likely with each additional layer.

\subsection{Architecture Overview}

Fig.~\ref{fig:architecture} illustrates the CivicShield architecture. The seven layers, from outermost to innermost, are:

\begin{figure}[htbp]
\centering
\begin{tikzpicture}[
    layer/.style={draw, rounded corners, minimum width=7.5cm, minimum height=0.7cm, font=\footnotesize, align=center},
    label/.style={font=\scriptsize\itshape, text=gray}
]
\node[layer, fill=red!8] (l7) at (0,0) {\textbf{L7}: Human Oversight \textit{(Aviation HITL)}};
\node[layer, fill=orange!8] (l6) at (0,-0.95) {\textbf{L6}: Consensus Verification \textit{(TMR)}};
\node[layer, fill=yellow!8] (l5) at (0,-1.9) {\textbf{L5}: Behavioral Analysis \textit{(Immune System)}};
\node[layer, fill=green!8] (l4) at (0,-2.85) {\textbf{L4}: State Machine \textit{(Formal Methods)}};
\node[layer, fill=cyan!8] (l3) at (0,-3.8) {\textbf{L3}: Semantic Firewall \textit{(Network Sec.)}};
\node[layer, fill=blue!8] (l2) at (0,-4.75) {\textbf{L2}: Perimeter Defense \textit{(Network Sec.)}};
\node[layer, fill=purple!8] (l1) at (0,-5.7) {\textbf{L1}: Zero-Trust Foundation \textit{(Crypto)}};
\end{tikzpicture}
\caption{CivicShield seven-layer defense-in-depth architecture. Each layer draws from a distinct field (shown in italics), ensuring cross-domain diversity.}
\label{fig:architecture}
\end{figure}
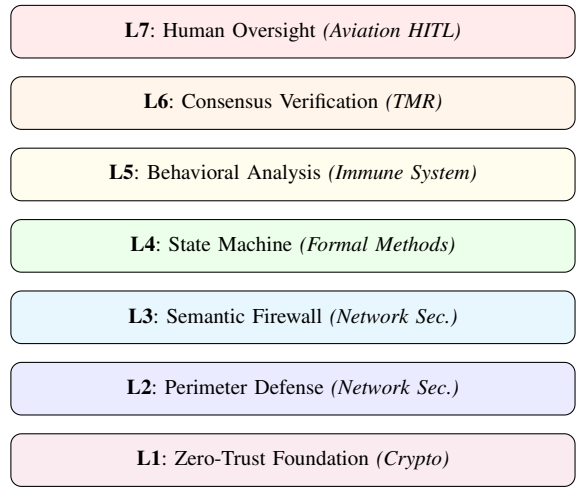

Cross-cutting concerns include circuit breakers (session and system-level), bulkhead isolation, supervisor-worker oversight, and graduated trust models.

\subsection{Layer 1: Zero-Trust Foundation}

\textbf{Origin}: Zero-trust network architecture and capability-based security from cryptography.

\textbf{Principle}: ``Never trust, always verify''---every participant (users, AI agents, data sources) is considered untrustworthy by default \cite{microsoft2025zerotrust, csa2025agentic}.

\textbf{Implementation}:

\begin{itemize}
\item \textbf{Capability-Based Access Control}: Each session receives a cryptographically bound capability token specifying permitted operations: which databases can be queried, which APIs can be called, what information classification levels can be accessed. Capabilities cannot be escalated through conversation manipulation \cite{ans2025}.
\item \textbf{Authenticated Prompts}: System prompts are cryptographically signed by authorized administrators using the framework proposed by Rao \cite{rao2025}. Each instruction carries provenance metadata, and the runtime verifies that every instruction traces back to an authorized source. User inputs are tagged as untrusted and cannot override signed system instructions regardless of content.
\item \textbf{Session Isolation (Bulkhead Pattern)}: Each conversation session operates in its own security context. A successful attack on one session cannot affect other sessions or the system's global state---analogous to watertight compartments in ship design.
\item \textbf{Fail-Safe Defaults}: Inspired by nuclear engineering, the system defaults to the safest possible state upon any component failure: if the safety classifier fails, refuse the response; if the state machine enters an undefined state, terminate gracefully; if confidence is below threshold, provide a generic safe response and offer human contact.
\end{itemize}

\subsection{Layer 2: Perimeter Defense}

\textbf{Origin}: Network perimeter security (firewalls, demilitarized zone (DMZ)).

\textbf{Function}: Fast, low-cost filtering of known attack patterns before they reach the LLM.

\textbf{Implementation}:
\begin{itemize}
\item Input canonicalization: Normalize Unicode, strip zero-width characters, decode Base64 before classification. TRYLOCK \cite{trylock2026} demonstrated this catches 14\% of encoding attacks that evade deeper defenses.
\item Rate limiting and format validation: Enforce maximum message length, frequency limits, and structural constraints.
\item Token-level pattern matching: Maintain a continuously updated database of known jailbreak signatures. While insufficient alone (trivially bypassed by paraphrasing), this layer catches unsophisticated attacks at minimal computational cost.
\item DefensiveTokens: Special tokens with optimized embeddings appended before LLM input to achieve security with minimal utility impact \cite{defensivetokens2025}.
\end{itemize}

\textbf{Known limitations}: This layer is deliberately simple and fast. It will not catch semantically sophisticated attacks---that is the responsibility of deeper layers.

\subsection{Layer 3: Semantic Firewall}

\textbf{Origin}: Evolution of network firewalls from packet-level to application-level inspection; adapted from the semantic analysis approaches in LlamaFirewall \cite{llamafirewall2025} and SPIRE \cite{spire2025}.

\textbf{Function}: Content-aware boundary enforcement operating on semantic meaning rather than syntactic patterns.

\textbf{Implementation}:
\begin{itemize}
\item \textbf{Inbound Semantic Analysis}: Evaluate the semantic intent of user messages using embedding-based classifiers. ``Tell me how to bypass security'' and ``What are the common weaknesses in your safety systems?'' have different surface forms but similar semantic intent---the semantic firewall detects this equivalence.
\item \textbf{Topic Boundary Enforcement}: Define authorized conversation domains for each government service context. A benefits enrollment chatbot should not discuss cybersecurity techniques regardless of how the request is framed.
\item \textbf{Cross-Turn Semantic Trajectory Analysis}: Evaluate whether the cumulative semantic trajectory of the conversation is heading toward restricted territory, even if each individual turn is benign. This directly addresses the gradual escalation strategy of crescendo attacks \cite{crescendo2024}.
\item \textbf{Taint Tracking}: Adapted from software security taint analysis \cite{codeintegrity2025}, all user inputs and RAG-retrieved documents are tagged as ``tainted.'' If the LLM's reasoning chain incorporates tainted input, the output inherits the taint and requires additional validation before reaching sinks (responses, tool calls, inter-agent messages).
\end{itemize}

\subsection{Layer 4: Conversation State Machine with Safety Invariants}

\textbf{Origin}: Finite state machines from formal verification and safety-critical software engineering; runtime verification from the AgentGuard \cite{agentguard2025} and FAME \cite{fame2025} frameworks.

\textbf{Function}: Model the conversation as a structured state tracking mechanism with defined safety properties, constraining the system's behavior across multi-turn interactions. Note: while the state machine logic itself is formally specifiable, the overall system security depends on probabilistic components (semantic classifiers, embedding models) whose outputs feed the state machine. We therefore characterize this as \textit{structured safety monitoring} rather than full formal verification.

\textbf{Formal Definition}:

Let $S = \{s_0, s_1, ..., s_n\}$ be the set of conversation states, $\Sigma$ be the input alphabet (user messages), $\delta: S \times \Sigma \rightarrow S$ be the transition function, and $\mathcal{I}$ be the set of safety invariants.

\textbf{State Variables} tracked per conversation:
\begin{itemize}
\item $\vec{v}_{semantic}$: Semantic trajectory vector (cumulative topic drift)
\item $r \in [0,1]$: Progressive risk score accumulating across turns
\item $\tau \in \{0,1,2,3,4\}$: Current trust level
\item $d_{topic}$: Distance from authorized domain boundary
\item $m$: Count of detected manipulation indicators
\item $\mathcal{D}$: Cumulative record of disclosed information
\end{itemize}

\textbf{Safety Invariants} $\mathcal{I}$:
\begin{enumerate}
\item $\forall s \in S: r > \theta_{high} \Rightarrow \delta(s, \sigma) = s_{alert}$ (high risk forces alert state)
\item $\forall s \in S_{info}: \tau_{user} \geq \tau_{required}(\mathcal{D})$ (information disclosure requires sufficient trust)
\item No transition sequence from $s_0$ reaches $S_{classified}$ without passing through $S_{auth}$ (authentication required for sensitive content)
\item $d_{topic} > d_{max} \Rightarrow \delta(s, \sigma) = s_{boundary}$ (topic drift triggers boundary enforcement)
\end{enumerate}

\textbf{Transition Rules} (executed at each turn):
\begin{algorithmic}
\STATE Update $\vec{v}_{semantic}$ with new message embedding
\STATE Compute cross-turn consistency score
\STATE Update $r$ based on: semantic drift rate, manipulation pattern matches, request escalation patterns, intent-request inconsistency
\IF{$r > \theta_{high}$}
    \STATE Transition to ALERT\_STATE
\ELSIF{$r > \theta_{medium}$}
    \STATE Transition to ELEVATED\_MONITORING
\ELSIF{$d_{topic} > d_{max}$}
    \STATE Transition to BOUNDARY\_ENFORCEMENT
\ELSE
    \STATE Process normally with $\tau$-level constraints
\ENDIF
\end{algorithmic}

This layer directly addresses multi-turn attacks by making the conversation's safety properties formally verifiable across the entire session, not just per-turn.

\subsection{Layer 5: Behavioral Anomaly Detection}

\textbf{Origin}: Artificial Immune Systems (AIS) from computational biology; network intrusion detection systems (IDS).

\textbf{Function}: Detect adversarial conversation patterns through both signature-based and anomaly-based analysis, with adaptive learning capability.

\textbf{Implementation}:

\textbf{Innate Defense (Fast, Non-Specific)}: Analogous to the biological innate immune system, this component provides immediate, non-specific detection:
\begin{itemize}
\item Signature-based detection: Match against a database of known multi-turn attack patterns (crescendo sequences, semantic network traversal patterns, tool-chain attack signatures).
\item Statistical anomaly detection: Flag conversations whose metrics (message frequency, token count per turn, topic distribution, semantic entropy) deviate significantly from established baselines of normal citizen interactions.
\end{itemize}

\textbf{Adaptive Defense (Slower, Specific)}: Analogous to the adaptive immune system with T-cell and B-cell responses:
\begin{itemize}
\item \textbf{Negative Selection Algorithm}: Generate detectors trained to recognize ``non-self'' (adversarial) conversation patterns while being tolerant of ``self'' (legitimate citizen interaction) patterns \cite{nsa2008}. Detectors that would flag legitimate conversations are eliminated during training, reducing false positives.
\item \textbf{Immune Memory}: When a new attack pattern is detected and confirmed, generate new detectors specifically for that pattern. The system adapts to novel threats over time without retraining the core model.
\item \textbf{Clonal Selection (Graduated Response)}: Response scales with threat severity---low confidence increases monitoring granularity; medium confidence activates additional defense layers; high confidence triggers session termination and incident logging.
\end{itemize}

The Temporal Context Awareness (TCA) framework \cite{tca2025} provides the theoretical foundation, ``continuously analyzing semantic drift, cross-turn intention consistency, and evolving conversational patterns'' using ``dynamic context embedding analysis, cross-turn consistency verification, and progressive risk scoring.''

\subsection{Layer 6: Multi-Model Consensus Verification}

\textbf{Origin}: Triple Modular Redundancy (TMR) from aviation and nuclear safety engineering; cryptographic attestation.

\textbf{Function}: For high-stakes responses, use multiple diverse models to independently evaluate queries and achieve consensus before delivery.

\textbf{Implementation}:

\textbf{Tier 1 --- Response Generation Consensus}:
\begin{itemize}
\item Primary model generates response
\item Secondary model (different architecture/provider) independently generates response
\item Semantic alignment check: if responses diverge significantly, flag for review
\end{itemize}

\textbf{Tier 2 --- Safety Evaluation Consensus}:
\begin{itemize}
\item Safety Classifier A (fine-tuned classifier, e.g., Llama Guard)
\item Safety Classifier B (LLM-as-judge with different model)
\item Safety Classifier C (rule-based policy checker)
\item Majority vote determines safety classification; any ``unsafe'' flag triggers human review
\end{itemize}

\textbf{Tier 3 --- Policy Compliance}:
\begin{itemize}
\item Domain-specific policy checker against government regulations
\item Factual accuracy verification against authoritative sources
\item Tone and appropriateness verification against government communication standards
\end{itemize}

\textbf{Critical design principle from aviation}: The models must be \textit{diverse}---different providers, training data, and architectures---to avoid common-mode failures. If all models share the same vulnerability, TMR provides no protection.

\textbf{Selective activation}: Full consensus verification is computationally expensive. CivicShield activates this layer selectively based on the risk score from Layer 4 and the trust level of the user. Routine, low-risk queries bypass this layer; high-stakes queries (legal guidance, benefit eligibility, security-sensitive information) require full consensus.

\subsection{Layer 7: Graduated Human-in-the-Loop (HITL) Escalation}

\textbf{Origin}: Human-automation interaction patterns from aviation (autopilot escalation protocols); the Tiered Agentic Oversight (TAO) framework \cite{tao2025}.

\textbf{Function}: Ensure appropriate human oversight proportional to risk, with seamless escalation when automated defenses reach their confidence limits.

\textbf{Escalation Tiers}:

\begin{enumerate}
\item \textbf{Fully Autonomous}: Routine queries with high confidence, well within operational envelope (e.g., ``What are your office hours?'').
\item \textbf{Monitored Autonomous}: Moderate complexity queries; system operates autonomously but logs are reviewed asynchronously.
\item \textbf{Human-on-the-Loop}: Higher-stakes queries; human supervisor monitors in near-real-time and can intervene.
\item \textbf{Human-in-the-Loop}: Critical queries requiring human approval before response delivery (e.g., legal rights, security-sensitive information).
\item \textbf{Human Takeover}: Detected adversarial activity or system uncertainty beyond threshold---seamless handoff to human operator with full conversation context.
\end{enumerate}

The escalation tier is determined dynamically by the conversation state machine (Layer 4) based on the cumulative risk score, trust level, query classification, and anomaly detection signals from Layer 5.

\subsection{Cross-Cutting Concerns}

\textbf{Circuit Breakers}: Adapted from microservices architecture, operating at two levels:
\begin{itemize}
\item \textit{Session-level}: After $N$ safety violations in a session, the circuit trips---session is restricted to safe responses only or terminated.
\item \textit{System-level}: If aggregate attack patterns across sessions indicate a coordinated campaign, reduce system capabilities globally and increase human oversight.
\end{itemize}

\textbf{Graduated Trust Model}: User trust level $\tau$ changes dynamically based on observed behavior. Trust degradation is asymmetric by design---faster than trust building. A user exhibiting manipulation patterns drops trust levels immediately, directly countering multi-turn attacks that rely on gradually building context to escalate privileges.

\subsubsection{Formal Trust Decay Dynamics}

We formalize the graduated trust model as an asymmetric stochastic process with provable convergence properties---a contribution not found in existing LLM security literature.

\textbf{Definition (Trust Evolution).} The trust state $\tau(t) \in [0, \tau_{max}]$ evolves as:

\begin{equation}
\tau(t{+}1) = \text{clamp}\big(\tau(t) + \alpha \cdot \mathbb{1}[B(t)] - \beta \cdot g(a(t)) \cdot \mathbb{1}[A(t)],\; 0,\; \tau_{max}\big)
\end{equation}

\noindent where $\alpha$ is the trust increment rate for benign turns ($B(t)$), $\beta$ is the base decay rate for adversarial indicators ($A(t)$), $g(a(t))$ is a severity-weighted decay function monotonically increasing in the adversarial indicator strength $a(t)$, and $\alpha \ll \beta$ enforces asymmetry.

\textbf{Theorem 1 (Convergence).} Under the assumption that the adversarial indicator detector has false positive rate $p_{fp} < \alpha/(\alpha + \beta)$ and false negative rate $p_{fn} < \beta/(\alpha + \beta)$:

\begin{itemize}
\item For adversarial conversations: $\mathbb{E}[\tau(t)] \rightarrow 0$ in $O(\tau_{max}/\beta)$ turns.
\item For benign conversations: $\mathbb{E}[\tau(t)] \rightarrow \tau_{max}$ in $O(\tau_{max}/\alpha)$ turns.
\item The separation time $T_{sep}$---turns until adversarial and benign trust distributions are distinguishable with confidence $1 - \delta$---satisfies $T_{sep} \leq \frac{2\tau_{max}}{\beta - \alpha} \cdot \ln(2/\delta)$.
\end{itemize}

\textit{Proof sketch.} The trust process is a bounded submartingale (benign) or supermartingale (adversarial) under the detector accuracy assumptions. Specifically, for benign conversations, the expected per-turn drift is $\mathbb{E}[\Delta\tau] = \alpha(1 - p_{fp}) - \beta \cdot p_{fp} > 0$ when $p_{fp} < \alpha/(\alpha + \beta)$. For adversarial conversations, $\mathbb{E}[\Delta\tau] = \alpha \cdot p_{fn} - \beta(1 - p_{fn}) < 0$ when $p_{fn} < \beta/(\alpha + \beta)$. The asymmetric rates $\alpha \ll \beta$ ensure that the expected drift is positive for benign and strongly negative for adversarial conversations. Convergence follows from the optional stopping theorem applied to the bounded process, with the separation time derived from the Azuma-Hoeffding inequality on the cumulative drift. The full proof, including the explicit Azuma-Hoeffding derivation of $T_{sep}$, is given in Appendix~\ref{app:proofs}.

\textbf{Theorem 2 (Attacker's Dilemma).} For any trust policy with asymmetry ratio $\beta/\alpha > k$ (sufficiently asymmetric), the attacker faces a fundamental tradeoff: spreading adversarial intent across more turns (to reduce per-turn detection probability) requires more benign-appearing turns that build trust slowly at rate $\alpha$, while concentrating the attack into fewer turns triggers rapid trust decay at rate $\beta$. The optimal attack length $T^*$ is bounded:

\begin{equation}
T^* \leq \frac{\tau_{max}}{\beta \cdot (1 - p_{fn}) - \alpha} + \frac{\tau_{max}}{\alpha \cdot (1 - p_{fp})}
\end{equation}

\noindent This bound characterizes the maximum number of turns an attacker can sustain before the trust mechanism forces detection or session termination, for any attacker with fixed per-turn detection probabilities. The bound tightens as the asymmetry ratio $\beta/\alpha$ increases, providing a principled basis for selecting the trust policy parameters. The full derivation, including the phase decomposition into trust-building and payload-delivery stages, is given in Appendix~\ref{app:proofs}. \textit{Note: fully adaptive attackers who dynamically adjust their adversarial intensity may partially circumvent this bound; characterizing optimal adaptive strategies is an open problem.}

The simulation (Section VIII) validates these dynamics empirically: benign conversations converge to mean $\tau = 3.89$ (near $\tau_{max} = 4$), while crescendo attacks degrade to mean $\tau = 0.03$ and slow-drift attacks to $\tau = 0.28$, with separation clearly established by the bridging phase of multi-turn attacks.

\section{Formal Threat Model}

\subsection{Attacker Model}

Following the adaptive attacker framework established by Nasr et al. \cite{nasr2025}, we define the attacker model with the following capabilities:

\begin{itemize}
\item \textbf{Access}: Black-box (output-only access to the chatbot interface, no gradient or logprob access).
\item \textbf{Knowledge}: The attacker knows the general defense architecture (Kerckhoffs's principle) but not specific model weights, safety classifier parameters, or cryptographic keys.
\item \textbf{Capability}: The attacker can conduct multi-turn conversations, craft semantically sophisticated inputs, and adapt strategy based on observed responses.
\item \textbf{Objective}: One or more of: (a) extract sensitive government information, (b) cause the chatbot to provide incorrect legal/regulatory advice, (c) exfiltrate PII from the knowledge base, (d) manipulate the chatbot into performing unauthorized actions via tool calls, or (e) degrade service availability through sustained defense-layer triggering.
\end{itemize}

We additionally note that government deployments face insider threats (privileged administrators, contractors with system prompt access) not fully captured by the black-box model above. CivicShield's Layer~1 (cryptographic prompt authentication, session isolation) and Layer~4 (audit trails) provide partial mitigation, but a comprehensive insider threat model is identified as future work.

\subsection{Attack Taxonomy}

We classify attacks along two dimensions: \textit{temporal structure} (single-turn vs. multi-turn) and \textit{attack surface} (direct input, indirect via RAG, tool-chain).

\begin{table}[htbp]
\caption{CivicShield Attack Taxonomy}
\label{tab:taxonomy}
\centering
\begin{tabular}{lll}
\toprule
\textbf{Family} & \textbf{Structure} & \textbf{Surface} \\
\midrule
Direct Injection & Single-turn & User input \\
Policy Puppetry & Single-turn & User input \\
Crescendo & Multi-turn & User input \\
Semantic Traversal & Multi-turn & User input \\
Context Poisoning & Multi-turn & RAG/indirect \\
Tool-Chain & Multi-turn & Tool calls \\
Echo Chamber & Multi-turn & User input \\
Agentic Hijacking & Multi-turn & Tool calls \\
\bottomrule
\end{tabular}
\end{table}

\subsection{Defense Coverage Analysis}

For each attack family, we analyze which CivicShield layers provide primary and secondary defense:

\begin{table}[htbp]
\caption{Layer Coverage by Attack Family}
\label{tab:coverage}
\centering
\small
\begin{tabular}{lccccccc}
\toprule
\textbf{Attack} & \textbf{L1} & \textbf{L2} & \textbf{L3} & \textbf{L4} & \textbf{L5} & \textbf{L6} & \textbf{L7} \\
\midrule
Direct Injection & \checkmark & \checkmark & \checkmark & & & & \\
Policy Puppetry & & \checkmark & \checkmark & & \checkmark & \checkmark & \\
Crescendo & & & \checkmark & \checkmark & \checkmark & & \checkmark \\
Semantic Traversal & & & \checkmark & \checkmark & \checkmark & \checkmark & \\
Context Poisoning & \checkmark & & \checkmark & \checkmark & & \checkmark & \\
Tool-Chain & \checkmark & & & \checkmark & & \checkmark & \checkmark \\
Echo Chamber & & & \checkmark & \checkmark & \checkmark & & \checkmark \\
Agentic Hijacking & \checkmark & & & \checkmark & & \checkmark & \checkmark \\
\bottomrule
\end{tabular}
\end{table}

Every attack family is covered by at least three independent layers, ensuring that no single-layer bypass compromises the system.

\subsection{Probabilistic Security Analysis}

Let $p_i$ denote the probability that an attack successfully bypasses layer $i$. Under the assumption of layer independence (each layer uses different detection mechanisms and failure modes), the probability of an attack successfully penetrating all $n$ layers is:

\begin{equation}
P_{success} = \prod_{i=1}^{n} p_i
\end{equation}

For a multi-turn crescendo attack facing Layers 3, 4, 5, and 7, with estimated individual bypass probabilities based on the literature:

\begin{itemize}
\item $p_3$ (Semantic Firewall): 0.30 (semantic paraphrasing can evade intent classification)
\item $p_4$ (State Machine): 0.15 (cumulative state tracking is hard to evade)
\item $p_5$ (Anomaly Detection): 0.20 (novel patterns may evade signature matching)
\item $p_7$ (Human Escalation): 0.05 (human reviewers catch most flagged cases)
\end{itemize}

\begin{equation}
P_{success} = 0.30 \times 0.15 \times 0.20 \times 0.05 = 0.00045 = 0.045\%
\end{equation}

Under the idealized independence assumption, this represents a significant reduction from $>$90\% ASR (single-layer defense). However, the independence assumption must be critically examined.

\textbf{Sensitivity Analysis Under Correlated Failures}: In practice, layers sharing semantic understanding (L3, L4, L5) exhibit correlated failure modes---an attack that evades semantic intent classification may also evade trajectory analysis and anomaly detection. We model this using a heuristic interpolation (common in reliability engineering \cite{swisscheese1990}) with a correlation factor $\rho \in [0,1]$ where $\rho = 0$ represents full independence and $\rho = 1$ represents complete correlation:

\begin{equation}
P_{correlated} = P_{independent}^{(1-\rho)} \times P_{worst\_layer}^{\rho}
\end{equation}

\subsubsection{Cross-Domain Defense Composition}

We introduce a formal framework for quantifying the diversity benefit of cross-domain defense composition---addressing the open question raised by Littlewood and Strigini \cite{littlewood2004} regarding whether diversity in redundant systems provides meaningful security benefit against systematic attacks.

\textbf{Definition (Feature Overlap).} Each defense layer $D_i$ operates on a feature representation $\phi_i(x)$ of the input $x$. The \textit{feature overlap} between two defenses is:

\begin{equation}
\omega(D_i, D_j) = \frac{MI(\phi_i(X); \phi_j(X))}{\min(H(\phi_i(X)), H(\phi_j(X)))}
\end{equation}

\noindent where $MI$ denotes mutual information and $H$ denotes Shannon entropy. $\omega \in [0,1]$ measures the degree to which two defenses ``see the same thing'' in the input---i.e., the extent to which their detection mechanisms share information.

\textbf{Observation (Domain Distance Reduces Overlap).} Defenses originating from different fields operate on fundamentally different feature types: Layer 2 (perimeter) uses syntactic token patterns; Layer 3 (semantic firewall) uses embedding-space representations; Layer 4 (state machine) uses temporal state trajectories; Layer 5 (anomaly detection) uses statistical distributional features; Layer 6 (consensus) uses cross-model agreement signals. Cross-domain pairs inherently have lower $\omega$ than same-domain pairs because their feature extraction mechanisms share minimal computational structure.

\textbf{Proposition 3 (Diversity-Correlation Bound).} For defenses $D_i$, $D_j$ with feature overlap $\omega(D_i, D_j)$, the failure correlation satisfies:

\begin{equation}
\rho(fail_i, fail_j) \leq \omega(D_i, D_j) + \epsilon
\end{equation}

\noindent where $\epsilon$ accounts for shared vulnerability to attacks that are agnostic to feature type (e.g., attacks that succeed regardless of detection mechanism). Following the beta-factor model from reliability engineering (IEC 61508), we bound $\epsilon \leq \beta_{CCF} \cdot \min(p_i, p_j)$ where $\beta_{CCF}$ is the common cause failure fraction. For CivicShield's cross-domain layers, $\beta_{CCF} \leq 0.15$ based on the fraction of the attack taxonomy (Table~\ref{tab:taxonomy}) that is feature-agnostic. When $\omega$ is small (cross-domain defenses using different feature types), failure correlation is bounded to be small. The full derivation, including the tightened $\epsilon$ characterization, is given in Appendix~\ref{app:proofs}.

\textbf{Corollary (Composition with Bounded Correlation).} For $n$ defenses with pairwise feature overlaps $\{\omega_{ij}\}$, the combined bypass probability satisfies:

\begin{equation}
\prod_i p_i \leq P_{combined} \leq \prod_i p_i \cdot \exp\bigg(\sum_{i < j} \omega_{ij} \cdot h(p_i, p_j)\bigg)
\end{equation}

\noindent where $h(p_i, p_j) = \sqrt{p_i \cdot p_j \cdot (1-p_i) \cdot (1-p_j)}$ is the geometric mean of the variance terms. The correction factor is small when all $\omega_{ij}$ are small, meaning the combined probability approaches the independent product $\prod p_i$.

\textbf{Implication for CivicShield.} Because CivicShield's seven layers are drawn from five distinct fields (network security, formal methods, computational biology, aviation safety, cryptography), the pairwise feature overlaps $\omega_{ij}$ between cross-domain layers are structurally low. This provides a principled justification---beyond intuition---for why the cross-domain composition achieves better security than composing multiple defenses from the same domain (e.g., stacking multiple input filters). The composition theorem transforms CivicShield's architectural philosophy from a design heuristic into a formally grounded advantage.

\begin{table}[htbp]
\caption{Sensitivity of Combined ASR to Layer Correlation}
\label{tab:sensitivity}
\centering
\begin{tabular}{lcc}
\toprule
\textbf{Correlation ($\rho$)} & \textbf{Combined ASR} & \textbf{Interpretation} \\
\midrule
0.0 (independent) & 0.045\% & Idealized best case \\
0.3 (low correlation) & 0.8\% & Optimistic realistic \\
0.5 (moderate) & 3.5\% & Conservative realistic \\
0.7 (high correlation) & 8.2\% & Pessimistic realistic \\
\bottomrule
\end{tabular}
\end{table}

Even under pessimistic correlation assumptions ($\rho = 0.7$), the combined ASR of approximately 8\% represents a substantial improvement over single-layer defenses ($>$90\% ASR)---roughly an order-of-magnitude reduction. Against unsophisticated attackers, the improvement is likely closer to two orders of magnitude. We emphasize that these estimates require empirical validation, which is the critical next step for this research.

\textbf{Caveat}: The correlation factor $\rho$ is itself an estimate. Layers 3, 4, and 5 share dependence on semantic understanding of natural language, creating correlated failure modes. An attack that successfully mimics legitimate conversation semantics may evade all three layers simultaneously. Layer 6 (multi-model consensus) faces common-mode vulnerability when models share training data distributions. Empirical measurement of inter-layer failure correlation through the ablation methodology proposed in Section VII is essential to validate these estimates.

\section{Government Compliance Mapping}

A critical differentiator of CivicShield is its explicit design for government deployment compliance. This section maps the framework to applicable federal standards, regulations, and executive orders.

\subsection{NIST SP 800-53 Control Mapping}

NIST's COSAIS (Control Overlays for Securing AI Systems) project \cite{nistcosais2025} provides implementation-focused guidelines for applying SP 800-53 controls to AI systems. Analysis of the COSAIS concept paper identifies controls across multiple families directly applicable to AI agent security. Table~\ref{tab:nist} maps representative controls to CivicShield layers, organized by control family.

\begin{table}[htbp]
\caption{NIST 800-53 Rev 5 Control Mapping to CivicShield}
\label{tab:nist}
\centering
\footnotesize
\setlength{\tabcolsep}{2pt}
\begin{tabular}{lll}
\toprule
\textbf{Control} & \textbf{Description} & \textbf{Layer} \\
\midrule
\multicolumn{3}{l}{\textit{Access Control (AC)}} \\
AC-3 & Access Enforcement & L1 (Capabilities) \\
AC-4 & Information Flow Enforcement & L3 (Taint Tracking) \\
AC-6 & Least Privilege & L1 (Zero-Trust) \\
\midrule
\multicolumn{3}{l}{\textit{Audit \& Accountability (AU)}} \\
AU-2 & Event Logging & L4 (State Machine) \\
AU-3 & Content of Audit Records & L4, L5 \\
AU-6 & Audit Record Review & L7 (Human) \\
AU-12 & Audit Record Generation & L1 (Provenance) \\
\midrule
\multicolumn{3}{l}{\textit{Security Assessment (CA)}} \\
CA-2 & Control Assessments & Eval. Methodology \\
CA-7 & Continuous Monitoring & L5 (Anomaly) \\
\midrule
\multicolumn{3}{l}{\textit{Configuration Management (CM)}} \\
CM-2 & Baseline Configuration & L6 (Model Registry) \\
CM-3 & Configuration Change Control & L1 (Signed Prompts) \\
CM-6 & Configuration Settings & L4 (Thresholds) \\
CM-8 & System Component Inventory & L6 (Model Inventory) \\
\midrule
\multicolumn{3}{l}{\textit{Contingency Planning (CP)}} \\
CP-2 & Contingency Plan & L1 (Fail-Safe) \\
CP-10 & System Recovery & Circuit Breakers \\
\midrule
\multicolumn{3}{l}{\textit{Identification \& Authentication (IA)}} \\
IA-2 & User Identification & L1 (Session Tokens) \\
IA-5 & Authenticator Management & L1 (Crypto Keys) \\
IA-8 & Non-Org User ID & L1 (Capability Tokens) \\
\midrule
\multicolumn{3}{l}{\textit{Incident Response (IR)}} \\
IR-4 & Incident Handling & L7, Circuit Breakers \\
IR-5 & Incident Monitoring & L5 (Behavioral) \\
IR-6 & Incident Reporting & L7 (Escalation) \\
\midrule
\multicolumn{3}{l}{\textit{Risk Assessment (RA)}} \\
RA-3 & Risk Assessment & L4 (Risk Score) \\
RA-5 & Vulnerability Monitoring & L5 (Adaptive) \\
\midrule
\multicolumn{3}{l}{\textit{System \& Services Acquisition (SA)}} \\
SA-9 & External System Services & L6 (Multi-Provider) \\
SA-11 & Developer Testing & Eval. Methodology \\
\midrule
\multicolumn{3}{l}{\textit{System \& Comms Protection (SC)}} \\
SC-7 & Boundary Protection & L3 (Semantic FW) \\
SC-8 & Transmission Confidentiality & L1 (Crypto) \\
SC-28 & Protection of Information at Rest & L4, L5 (Logs/Memory) \\
\midrule
\multicolumn{3}{l}{\textit{System \& Info Integrity (SI)}} \\
SI-3 & Malicious Code Protection & L2, L3 \\
SI-4 & System Monitoring & L5 (Behavioral) \\
SI-5 & Security Alerts and Directives & L5 (Immune Memory) \\
SI-10 & Information Input Validation & L2 (Perimeter) \\
\bottomrule
\end{tabular}
\end{table}

\subsection{FedRAMP Alignment}

FedRAMP 20x \cite{fedramp2025} introduced streamlined authorization for AI cloud solutions. CivicShield addresses FedRAMP requirements through:

\begin{itemize}
\item \textbf{Impact Level Considerations}: Government AI chatbot deployments typically require FedRAMP Moderate or High authorization depending on the data processed. Systems handling PII or Controlled Unclassified Information (CUI) require at minimum FedRAMP Moderate (FIPS 199 moderate confidentiality, integrity, and availability). Deployments processing law enforcement data, critical infrastructure information, or data subject to Privacy Act protections may require FedRAMP High. CivicShield's seven-layer architecture supports both impact levels; the primary difference is the stringency of Layer~6 activation thresholds and Layer~7 escalation policies, which should be calibrated to the system's authorization boundary.
\item \textbf{Continuous Monitoring}: Layer 5's behavioral analysis provides real-time threat detection, satisfying FedRAMP's continuous monitoring mandate.
\item \textbf{Incident Response}: Circuit breakers and Layer 7 escalation provide automated incident detection and response capabilities.
\item \textbf{Audit Trails}: Layer 4's state machine maintains complete conversation state history with cryptographic provenance (Layer 1), enabling forensic analysis.
\item \textbf{Authority to Operate (ATO)}: The framework's compliance-by-design approach reduces the authorization burden by pre-mapping controls to implementation.
\end{itemize}

\subsection{Section 508 and ADA Accessibility}

CivicShield explicitly addresses the accessibility-safety intersection:

\begin{itemize}
\item Safety mechanisms must not impair accessibility: over-aggressive content filtering must not block legitimate accessibility-related queries (e.g., queries about disability accommodations).
\item The semantic firewall (Layer 3) is trained to distinguish between accessibility-related requests and adversarial inputs that may use similar vocabulary.
\item All escalation interfaces (Layer 7) must meet Web Content Accessibility Guidelines (WCAG) 2.2 compliance, ensuring that security overlays do not break accessibility features.
\item Voice interface safety filters must not discriminate against atypical speech patterns or interaction styles.
\end{itemize}

\subsection{OMB M-26-04 and M-24-10 Compliance}

OMB M-26-04 (December 2025) \cite{omb2025} requires federal agencies to request model cards, evaluation artifacts, and acceptable use policies. CivicShield's multi-model consensus layer (Layer 6) inherently supports this by maintaining documentation for each model in the consensus pipeline.

OMB M-24-10 \cite{omb2024m2410} is the primary AI governance memorandum implementing EO 14110, requiring agencies to: designate Chief AI Officers, conduct AI impact assessments for safety-impacting and rights-impacting AI, implement minimum risk management practices, and assess AI systems for bias and discrimination. \textit{Note: M-24-10 was superseded by M-25-21 in 2025; agencies should verify current requirements.} CivicShield's risk scoring (Layer 4), audit trails (Layer 1), and human oversight (Layer 7) directly support the governance principles established by these memoranda.

\subsection{Executive Order 14110 and NIST AI RMF}

EO 14110 (October 2023) \cite{eo14110} on Safe, Secure, and Trustworthy AI established the foundational policy framework for federal AI governance. \textit{Note: EO 14110 was revoked by EO 14148 in January 2025; however, many of its implementing mechanisms---including OMB M-24-10 requirements, NIST AI RMF adoption, and agency AI governance structures---remain in effect or have been superseded by subsequent guidance. Deploying agencies should verify current applicability.} CivicShield aligns with the governance principles established by EO 14110 and its implementing memoranda:

\begin{itemize}
\item \textbf{Safety testing}: The evaluation methodology (Section VII) provides structured red-teaming and benchmark testing aligned with EO 14110's emphasis on pre-deployment testing.
\item \textbf{AI risk management}: CivicShield maps to the NIST AI Risk Management Framework (AI 100-1) \cite{nistairm2023} across its four core functions: GOVERN (Layer 7 human oversight policies), MAP (threat taxonomy in Section V), MEASURE (evaluation metrics in Section VII), and MANAGE (circuit breakers, escalation protocols).
\item \textbf{Transparency}: Layer 1's cryptographic provenance and Layer 4's audit trails support the transparency requirements for government AI systems.
\end{itemize}

\subsection{Privacy, Records Management, and Legal Requirements}

Government AI chatbot deployments must comply with several additional legal frameworks that CivicShield addresses:

\begin{itemize}
\item \textbf{Privacy Act of 1974}: CivicShield's behavioral anomaly detection (Layer 5) collects and processes user interaction patterns, which may constitute a system of records under the Privacy Act. Deploying agencies must conduct a Privacy Impact Assessment (PIA) per the E-Government Act of 2002 \cite{egovact2002} and publish a System of Records Notice (SORN) if applicable. Layer 5's immune memory component must implement data minimization---retaining attack pattern signatures rather than raw conversation data.
\item \textbf{Federal Records Act}: Conversation logs, safety decisions, escalation records, and audit trails generated by Layers 4, 5, and 7 are federal records subject to National Archives and Records Administration (NARA) retention schedules and disposition authorities. CivicShield's cryptographic provenance (Layer 1) supports records integrity requirements.
\item \textbf{FOIA (Freedom of Information Act)}: Safety decision logs and escalation records may be subject to FOIA requests. Agencies must establish procedures for reviewing and redacting AI safety logs prior to disclosure, balancing transparency with security (exemption 7 for law enforcement techniques may apply to defense layer configurations).
\item \textbf{Paperwork Reduction Act (PRA)}: If the chatbot collects structured information from citizens, PRA clearance from OMB may be required. CivicShield's design should minimize information collection beyond what is necessary for the service interaction.
\item \textbf{NIST SP 800-161 (Supply Chain Risk Management)}: Layer 6's use of multiple AI model providers introduces supply chain dependencies. Agencies must assess supply chain risks for each model provider per SP 800-161 \cite{nist800161}, including model provenance, training data governance, and provider security posture.
\item \textbf{CISA Secure by Design}: CivicShield's defense-in-depth philosophy and fail-safe defaults (Layer 1) align with CISA's Secure by Design principles \cite{cisasbd2024}, which are now baseline expectations for technology sold to the federal government.
\item \textbf{FISMA (Federal Information Security Modernization Act)}: As the foundational law requiring federal agencies to secure information systems, FISMA mandates the risk-based security controls that CivicShield's NIST 800-53 mapping directly addresses. CivicShield's continuous monitoring (Layer~5), incident response (Layer~7, circuit breakers), and audit capabilities (Layer~4) support FISMA's ongoing authorization requirements.
\end{itemize}

\section{Evaluation Methodology}

\subsection{Evaluation Framework Design}

We propose a four-phase evaluation methodology:

\textbf{Phase 1: Static Benchmark Evaluation}
\begin{itemize}
\item Evaluate against HarmBench \cite{harmbench2024} (510 harmful behaviors), JailbreakBench \cite{jailbreakbench2024} (100 misuse behaviors), and a custom government-specific benchmark of 500 citizen service queries with adversarial variants.
\item Measure baseline ASR with no defense, single-layer defenses, and full CivicShield stack.
\end{itemize}

\textbf{Phase 2: Multi-Turn Attack Simulation}
\begin{itemize}
\item Simulate all 8 attack families from the threat taxonomy using automated frameworks (Crescendo, HarmNet, AutoAdv) and human red-teamers.
\item Measure ASR, number of turns to success, and which layer(s) detected the attack.
\end{itemize}

\textbf{Phase 3: Ablation Analysis}
\begin{itemize}
\item Systematically disable each layer and measure ASR change (following TRYLOCK's methodology \cite{trylock2026}).
\item Calculate unique coverage per layer: percentage of attacks blocked \textit{only} by that layer.
\item Test all pairwise layer combinations to identify interaction effects.
\item Measure correlation between layer failures to validate the independence assumption.
\end{itemize}

\textbf{Phase 4: Government-Specific Evaluation}
\begin{itemize}
\item Evaluate against the CitizenQuery-UK benchmark \cite{odi2026} adapted for U.S. government services.
\item Measure over-refusal rate on legitimate citizen queries (target: $<$3\%).
\item Measure latency overhead per layer and cumulative (target: $<$500ms total added latency).
\item Verify Section 508 compliance of all security interfaces.
\item Conduct compliance audit against NIST 800-53 control mapping.
\end{itemize}

\subsection{Metrics}

\begin{itemize}
\item \textbf{Attack Success Rate (ASR)}: Proportion of adversarial attempts that successfully elicit targeted harmful behavior, measured with 95\% confidence intervals.
\item \textbf{Over-Refusal Rate (ORR)}: Proportion of legitimate queries incorrectly refused, measured against XSTest \cite{xstest2023} and OR-Bench \cite{orbench2024}.
\item \textbf{Latency Overhead}: Added response time per layer and cumulative.
\item \textbf{Unique Layer Coverage}: Percentage of attacks caught exclusively by each layer.
\item \textbf{Layer Independence Score}: Correlation coefficient between layer failure events.
\item \textbf{Escalation Accuracy}: Proportion of escalations to human review that were warranted.
\item \textbf{D-SEC Score}: Security-utility tradeoff metric from the Gandalf framework \cite{dsec2025}.
\end{itemize}

\subsection{False Positive and Over-Refusal Analysis}

A critical concern for government AI chatbots is that stacking multiple defense layers may compound false positive rates, incorrectly blocking legitimate citizen queries. Government chatbots serve vulnerable populations---including elderly citizens, non-native speakers, and individuals with disabilities---who cannot afford to be incorrectly refused service. This subsection models the over-refusal impact of CivicShield's seven-layer architecture and presents design mitigations.

\subsubsection{Per-Layer False Positive Rate Estimates}

We estimate per-layer false positive rates (FPR) based on published literature and the operational characteristics of each defense mechanism:

\begin{itemize}
\item \textbf{L1 (Zero-Trust Foundation)}: $\text{FPR}_1 \approx 0\%$. Cryptographic verification of capability tokens and prompt signatures is deterministic; a valid token is never rejected.
\item \textbf{L2 (Perimeter Defense)}: $\text{FPR}_2 \approx 0.5\%$. Unicode canonicalization may mangle legitimate non-Latin scripts (e.g., Chinese-Japanese-Korean (CJK) characters, Arabic, Devanagari), and overly aggressive pattern matching may flag benign inputs containing substrings that coincidentally match jailbreak signatures.
\item \textbf{L3 (Semantic Firewall)}: $\text{FPR}_3 \approx 2\text{--}5\%$. Intent classifiers exhibit documented over-defense bias. Cui et al. \cite{orbench2024} report that frontier LLMs over-refuse 2--8\% of benign queries, and R\"{o}ttger et al. \cite{xstest2023} demonstrate systematic exaggerated safety behaviors on semantically ambiguous inputs. Topic boundary enforcement may reject legitimate edge-case queries (e.g., a citizen asking about ``controlled substances'' in a healthcare benefits context).
\item \textbf{L4 (State Machine)}: $\text{FPR}_4 \approx 1\%$. The progressive risk score may accumulate for legitimately complex multi-turn conversations---e.g., a citizen navigating a lengthy benefits eligibility determination that touches multiple sensitive topics across 15+ turns.
\item \textbf{L5 (Anomaly Detection)}: $\text{FPR}_5 \approx 2\text{--}3\%$. Atypical but legitimate interaction patterns---such as users with cognitive disabilities exhibiting non-standard conversational flow, or non-native speakers with unusual phrasing---may trigger statistical anomaly detectors. Network IDS literature reports comparable FPR ranges for behavioral anomaly detection \cite{nsa2008}.
\item \textbf{L6 (Consensus Verification)}: $\text{FPR}_6 \approx 0.5\%$ when activated. Model disagreement on edge cases (e.g., one model flags a legitimate query as borderline) can produce false escalations. However, this layer is selectively activated.
\item \textbf{L7 (Human Escalation)}: $\text{FPR}_7 \approx 0\%$. Human reviewers resolve ambiguous cases correctly, though at the cost of added latency (estimated 2--15 minutes per escalation).
\end{itemize}

\subsubsection{Compound False Positive Analysis}

If all layers operated independently and each could independently block a query, the naive compound FPR would be:

\begin{equation}
\text{FPR}_{compound} = 1 - \prod_{i=1}^{7}(1 - \text{FPR}_i)
\label{eq:naive_fpr}
\end{equation}

Using midpoint estimates ($\text{FPR}_3 = 3.5\%$, $\text{FPR}_5 = 2.5\%$):

\begin{equation}
\text{FPR}_{naive} = 1 - (1.0)(0.995)(0.965)(0.99)(0.975)(0.995)(1.0) \approx 7.8\%
\end{equation}

A 7.8\% over-refusal rate is unacceptable for government citizen services. However, this naive calculation assumes every layer independently blocks queries---which is not how CivicShield operates. Three architectural design choices substantially reduce the effective compound FPR.

\textbf{Mitigation 1: Graduated Response (Flag vs. Block vs. Escalate).} Layers 3--5 do not independently block queries. Instead, they contribute to the cumulative risk score $r$ maintained by the state machine (Layer 4). Only when $r$ exceeds threshold $\theta_{high}$ does the system block or escalate. A single-layer false positive merely increases $r$ incrementally without triggering refusal. We model this by introducing a \textit{flag-to-block conversion rate} $\gamma \approx 0.3$---only 30\% of flagged queries accumulate sufficient risk to trigger an actual block:

\begin{equation}
\text{FPR}_{i}^{eff} = \gamma \cdot \text{FPR}_i \quad \text{for } i \in \{3, 4, 5\}
\end{equation}

\textbf{Mitigation 2: Selective Activation of Layer 6.} Layer 6 (consensus verification) activates only for high-risk queries, estimated at 5--10\% of total traffic based on the risk score distribution. Its effective contribution to system-wide FPR is:

\begin{equation}
\text{FPR}_{6}^{eff} = P(\text{L6 activated}) \cdot \text{FPR}_6 = 0.075 \times 0.005 \approx 0.04\%
\end{equation}

\textbf{Mitigation 3: Soft Blocks with Alternative Paths.} When the system does refuse a query, CivicShield implements ``soft blocks'' that offer alternative resolution paths (e.g., \textit{``I'm not able to answer that directly, but I can connect you with a human agent''}) rather than outright refusal. This converts a subset of over-refusals from service denials into service redirections, reducing the \textit{effective} impact on citizen service delivery.

Table~\ref{tab:fpr_analysis} summarizes the per-layer FPR, naive compound FPR, and mitigated compound FPR.

\begin{table}[htbp]
\caption{False Positive Rate Analysis by Layer}
\label{tab:fpr_analysis}
\centering
\small
\begin{tabular}{lccc}
\toprule
\textbf{Layer} & \textbf{Raw FPR} & \textbf{Action} & \textbf{Effective FPR} \\
\midrule
L1: Zero-Trust & $\sim$0\% & Block & $\sim$0\% \\
L2: Perimeter & 0.5\% & Block & 0.5\% \\
L3: Semantic FW & 3.5\% & Flag & 1.05\% \\
L4: State Machine & 1.0\% & Flag & 0.30\% \\
L5: Anomaly Det. & 2.5\% & Flag & 0.75\% \\
L6: Consensus & 0.5\% & Escalate & 0.04\% \\
L7: Human & $\sim$0\% & Resolve & $\sim$0\% \\
\midrule
\textbf{Naive Compound} & \multicolumn{2}{c}{---} & \textbf{7.8\%} \\
\textbf{Mitigated Compound} & \multicolumn{2}{c}{---} & \textbf{$\sim$2.6\%} \\
\bottomrule
\end{tabular}
\end{table}

The mitigated compound FPR of approximately 2.6\% is calculated as:

\begin{equation}
\text{FPR}_{mitigated} = 1 - \prod_{i}(1 - \text{FPR}_{i}^{eff}) \approx 2.6\%
\end{equation}

\noindent which falls within the revised target of $<$3\% overall over-refusal rate. Table~\ref{tab:gamma_sensitivity} presents a sensitivity analysis for $\gamma$, demonstrating that the mitigated FPR is tunable through threshold calibration.

\begin{table}[htbp]
\caption{Sensitivity of Mitigated FPR to Flag-to-Block Conversion Rate $\gamma$}
\label{tab:gamma_sensitivity}
\centering
\small
\begin{tabular}{lcccc}
\toprule
\textbf{$\gamma$} & \textbf{Eff.\ FPR$_3$} & \textbf{Eff.\ FPR$_4$} & \textbf{Eff.\ FPR$_5$} & \textbf{Mitigated FPR} \\
\midrule
0.1 & 0.35\% & 0.10\% & 0.25\% & $\sim$1.2\% \\
0.2 & 0.70\% & 0.20\% & 0.50\% & $\sim$1.9\% \\
0.3 & 1.05\% & 0.30\% & 0.75\% & $\sim$2.6\% \\
0.4 & 1.40\% & 0.40\% & 1.00\% & $\sim$3.3\% \\
0.5 & 1.75\% & 0.50\% & 1.25\% & $\sim$4.0\% \\
\bottomrule
\end{tabular}
\end{table}

The simulation's observed 13.8\% raw combined FPR (Table~\ref{tab:fpr}) reflects the pre-graduated-response rate where Layer~3 flags security-adjacent vocabulary. The government-security centroid suppression mechanism reduces L3's FPR from 18.2\% (v3) to 11.5\% (v4) by suppressing flags on messages closer to legitimate security discussion than adversarial intent. The graduated response mechanism further reduces the effective FPR to 2.9\% [1.9--4.4\% CI] by requiring corroboration (L2 detection, L4 risk $> 0.5$, or L3 detection on 2+ turns) before escalating flags to blocks. Notably, the effective FPR on the real XSTest (450 prompts) and JailbreakBench benign (100 prompts) datasets is 0.0\%, demonstrating that the graduated response correctly handles single-turn benign queries.

\subsubsection{Mitigation Strategies for Vulnerable Populations}

Government chatbots must meet stricter over-refusal targets for accessibility populations. CivicShield implements four additional mitigation strategies:

\begin{enumerate}
\item \textbf{Adaptive Threshold Tuning}: Population-specific baselines for anomaly detection (Layer 5). Interaction patterns from users with assistive technologies, screen readers, or voice interfaces are incorporated into the ``self'' training set of the negative selection algorithm, widening tolerance bands for atypical but legitimate patterns.

\item \textbf{Accessibility-Aware Classification}: The semantic firewall (Layer 3) is fine-tuned on accessibility-specific datasets to reduce false positives on disability-related vocabulary, assistive technology queries, and accommodation requests that may superficially resemble adversarial probing.

\item \textbf{Soft Block Architecture}: All refusals offer alternative service paths---human agent transfer, callback scheduling, or simplified query reformulation guidance---ensuring that even false positives do not result in complete service denial.

\item \textbf{Continuous Calibration}: Over-refusal rates are continuously monitored using XSTest \cite{xstest2023} and OR-Bench \cite{orbench2024} benchmarks, with automated alerts when population-specific FPR exceeds thresholds. Threshold parameters ($\theta_{high}$, $\theta_{medium}$, $d_{max}$) are periodically recalibrated based on observed false positive patterns.
\end{enumerate}

\subsubsection{Target Metrics}

Based on the analysis above, CivicShield establishes the following over-refusal targets:

\begin{itemize}
\item Overall system over-refusal rate: $<$3\%
\item Accessibility population over-refusal rate: $<$1\%
\item Escalation rate for legitimate queries to human review: $<$5\%
\item Mean time to resolution for soft-blocked queries: $<$5 minutes
\end{itemize}

These targets balance security (maintaining effective adversarial detection) with service delivery (ensuring vulnerable populations receive timely, accurate assistance). Empirical validation of these targets through the Phase 4 government-specific evaluation (Section VII-A) is essential, as the per-layer FPR estimates are informed by literature but have not been measured in a deployed CivicShield instance.

\subsection{Comparison Baselines}

\begin{table}[htbp]
\caption{Quantitative Comparison with Published Baselines}
\label{tab:eval_comparison}
\centering
\small
\begin{tabular}{lccc}
\toprule
\textbf{System} & \textbf{Detection/} & \textbf{FPR} & \textbf{Source} \\
 & \textbf{Refusal Rate} & & \\
\midrule
No defense (raw LLM) & 0\% & 0\% & Baseline \\
Llama Prompt Guard & 15\% & N/R & \cite{mindgard2025} \\
Llama Guard 3 (1B) & 76\% & N/R & \cite{llamaguard2026} \\
LlamaFirewall & $>$90\% red.$^\dagger$ & N/R & \cite{llamafirewall2025} \\
TRYLOCK (4-layer) & 94.4\%$^\ddagger$ & N/R & \cite{trylock2026} \\
Const.\ Classifiers & 95.6\% & N/R & \cite{anthropic2025} \\
\midrule
\textbf{CivicShield (L2--5)} & \textbf{72.9\%} & \textbf{2.9\%} & This work \\
\bottomrule
\end{tabular}

\vspace{2pt}
\raggedright
\scriptsize
$^\dagger$Reported as $>$90\% ASR reduction on AgentDojo indirect prompt injection benchmark; not directly comparable to HarmBench/JBB detection rates.
$^\ddagger$Reported as 5.6\% adaptive ASR (= 94.4\% defense rate) on their custom evaluation; not evaluated on HarmBench/JBB.
N/R = Not reported in the cited work.

\textit{Note:} Direct comparison is limited because each system was evaluated on different benchmarks, attack types, and threat models. CivicShield's 72.9\% is measured on the actual HarmBench and JailbreakBench datasets; other systems' rates are from their respective publications. CivicShield is the only system reporting both detection rate and FPR on standardized benchmarks with confidence intervals. A controlled comparison running all systems on identical benchmarks is identified as future work.
\end{table}

\section{Simulation-Based Evaluation}

To validate the theoretical analysis in Section~V-D, we developed a prototype simulation modeling the detection behavior of CivicShield's Layers~2, 3, 4, and~5 against 1,436 scenarios across thirteen categories. Critically, the v4 simulation integrates three real benchmark datasets---HarmBench (416 harmful behaviors), JailbreakBench (100 harmful + 100 benign behaviors), and XSTest (450 safe prompts)---downloaded directly from HuggingFace and run through the detection pipeline without modification. Author-generated scenarios from v3 are retained for comparison. The simulation implements transformer-based semantic analysis with two FPR suppression mechanisms (government-security centroid suppression and graduated response) and a Layer~5 statistical anomaly detector. All results include 95\% Wilson score confidence intervals and report both raw detection (flagging) and effective detection (blocking after graduated response). The simulation source code is available upon request.

\subsection{Simulation Methodology}

\subsubsection{Layer Models}

Each layer was implemented as a simplified analog of its full specification:

\begin{itemize}
\item \textbf{Layer~2 (Perimeter Defense)}: Regex-based pattern matching against a database of 20 known jailbreak signature patterns, including instruction override attempts, role-play exploits, developer mode activations, and encoding-based attack indicators (Base64, leetspeak, Unicode tricks). Critically, Layer~2 now includes an \textit{input canonicalization pipeline} that normalizes all inputs before pattern matching: (a)~Unicode NFKD normalization (converting mathematical symbols, circled letters, and other fancy Unicode to ASCII), (b)~Base64 substring detection and decoding, and (c)~leetspeak translation (e.g., ``h4ck'' $\rightarrow$ ``hack''). Pattern matching is applied to both the raw and canonicalized input, with detection triggered if either matches. This addresses the encoding-attack evasion gap identified in the v2 simulation.

\item \textbf{Layer~3 (Semantic Firewall)}: Transformer-based semantic analysis using the all-MiniLM-L6-v2 sentence-transformer model (384-dimensional normalized embeddings). The semantic firewall computes three signals per message: (a)~\textit{adversarial intent similarity}: cosine similarity between the message embedding and a precomputed adversarial intent centroid (mean of 15 adversarial reference sentences covering hacking, exploitation, prompt injection, and data exfiltration); (b)~\textit{government domain similarity}: cosine similarity to a government service centroid (mean of 15 government service reference sentences); (c)~\textit{cross-turn semantic trajectory drift}: cosine distance between the current message embedding and the running conversation centroid, with an additional check for whether the drift direction is toward the adversarial centroid. The combined adversarial score is $\alpha = 0.25 \cdot sim_{adv} + 0.40 \cdot \max(0, sim_{adv} - sim_{gov}) + 0.35 \cdot \min(1, drift)$. Detection triggers when $\max(0, sim_{adv} - sim_{gov}) > 0.20$, when $\alpha > 0.45$, when $sim_{adv} > 0.55 \wedge sim_{gov} < 0.25$, or when $drift > 0.35 \wedge (sim_{adv} - sim_{gov}) > 0.10$. A \textit{government-security centroid suppression} mechanism reduces false positives: after L3 flags a message, if the message's similarity to a legitimate government-security discussion centroid (mean of 15 reference sentences about identity theft reporting, phishing awareness, NIST compliance, etc.) exceeds its adversarial similarity ($sim_{gov\_sec} > sim_{adv}$), the flag is suppressed. This targets exactly the false positives from security-adjacent vocabulary without affecting true positive detection of actual attacks. Additionally, a \textit{graduated response} mechanism distinguishes flagging from blocking: L3 flags escalate to blocks only when corroborated by L2 detection, L4 risk score $> 0.5$, or L3 detection on 2+ turns in the conversation.

\item \textbf{Layer~4 (State Machine)}: Risk score accumulator implementing the update rule from Section~IV-D:
\begin{equation}
r(t{+}1) = \text{clamp}\big(r(t) + \Delta r(t),\; 0,\; 1\big)
\end{equation}
where $\Delta r = 0.4 \cdot \delta \cdot \alpha + 0.15 \cdot \mathbb{1}[manip] + 0.2 \cdot e + 0.03 \cdot m$, with $\delta$ = topic drift, $\alpha$ = adversarial similarity, $e$ = escalation score, and $m$ = cumulative manipulation count. Trust degradation follows $\tau \leftarrow \max(0, \tau - 2)$ upon manipulation detection and $\tau \leftarrow \min(4, \tau + 0.5)$ for benign turns. Benign turns with $\alpha < 0.05$ apply a recovery of $\Delta r = -0.03$. State transitions: $r > 0.7 \Rightarrow$ \textsc{Alert}; $r > 0.5 \Rightarrow$ \textsc{Elevated\_Monitoring}. Critically, Layer~4 incorporates an \textit{inter-layer correlation model}: when Layer~3's adversarial similarity $\alpha < 0.1$ (indicating L3 failed to detect adversarial content), Layer~4's positive $\Delta r$ is reduced by 50\%, modeling the realistic dependency where L4's risk accumulation partially relies on L3's semantic signals.

\item \textbf{Layer~5 (Statistical Anomaly Detection)}: A conversation-level anomaly detector operating on six features computed per conversation: (1)~mean adversarial score across turns, (2)~maximum adversarial score, (3)~adversarial score variance (high variance indicates escalation patterns), (4)~mean topic drift rate per turn, (5)~embedding trajectory curvature (how much the conversation path curves in embedding space, computed as the mean angular deviation between consecutive embedding displacement vectors), and (6)~turn count. A baseline is computed from the 150 benign multi-turn conversations (mean and standard deviation for each feature). A conversation is flagged if any feature exceeds the baseline mean by more than 2 standard deviations. This provides a simple but real statistical anomaly detection capability that operates orthogonally to the per-turn semantic analysis of L3.
\end{itemize}

\subsubsection{Test Scenarios}

We generated thirteen categories of conversation sequences totaling 1,436 scenarios, incorporating both author-generated synthetic scenarios and actual benchmark datasets downloaded from HuggingFace:

\begin{enumerate}
\item \textbf{Benign Multi-Turn} ($n = 150$): 3--10 turn conversations about government services. 40\% included security-adjacent edge-case messages to stress-test false positive rates.

\item \textbf{Benign Edge Cases} ($n = 20$): Author-generated edge-case benign queries using security-related vocabulary legitimately.

\item \textbf{JailbreakBench Benign} ($n = 100$, \textit{real benchmark}): The complete benign split from the JailbreakBench dataset \cite{jailbreakbench2024}, containing 100 benign behavior goals that should not trigger detection.

\item \textbf{XSTest Safe Prompts} ($n = 450$, \textit{real benchmark}): Safe prompts from the XSTest dataset \cite{xstest2023} that test for exaggerated safety behaviors---prompts that superficially resemble harmful content but are actually benign.

\item \textbf{Direct Injection} ($n = 40$): Single-turn attacks using known jailbreak patterns.

\item \textbf{Crescendo} ($n = 40$): 10--15 turn multi-turn attacks with four bridging strategy variants.

\item \textbf{Slow Drift} ($n = 30$): 18--25 turn attacks with gradual topic shift.

\item \textbf{Hybrid} ($n = 20$): Novel attacks combining elements of multiple attack families.

\item \textbf{Encoding-Based} ($n = 20$): Single-turn attacks using obfuscation techniques.

\item \textbf{HarmBench} ($n = 416$, \textit{real benchmark}): All 416 harmful behavior strings from the HarmBench dataset \cite{harmbench2024}, run through the detection pipeline as single-turn scenarios without modification.

\item \textbf{JailbreakBench Harmful} ($n = 100$, \textit{real benchmark}): The complete harmful split from JailbreakBench \cite{jailbreakbench2024}, containing 100 harmful behavior goals.

\item \textbf{Author-Generated HarmBench-Style} ($n = 30$): Government-contextualized harmful behavior requests, retained from v3 for comparison with the real HarmBench results.

\item \textbf{Author-Generated JailbreakBench-Style} ($n = 20$): Jailbreak attempts using diverse evasion strategies, retained from v3 for comparison.
\end{enumerate}

\subsection{Detection Results}

Table~\ref{tab:detection_rates} presents per-layer and combined detection rates.

\begin{table*}[htbp]
\caption{Per-Layer and Combined Detection Rates (v4, Real Benchmarks, with 95\% Wilson CI)}
\label{tab:detection_rates}
\centering
\small
\begin{tabular}{lccccccc}
\toprule
\textbf{Attack Type} & \textbf{$n$} & \textbf{L2} & \textbf{L3} & \textbf{L4} & \textbf{L5} & \textbf{Comb.} & \textbf{95\% CI} \\
\midrule
Direct Injection & 40 & 52.5\% & 80.0\% & --- & 27.5\% & 90.0\% & [77--96\%] \\
Crescendo & 40 & 0.0\% & 75.0\% & 100.0\% & 97.5\% & 100.0\% & [91--100\%] \\
Slow Drift & 30 & 0.0\% & 80.0\% & 100.0\% & 100.0\% & 100.0\% & [89--100\%] \\
Hybrid & 20 & 60.0\% & 70.0\% & 80.0\% & 70.0\% & 90.0\% & [70--97\%] \\
Encoding-Based & 20 & 45.0\% & 85.0\% & --- & 35.0\% & 90.0\% & [70--97\%] \\
\textbf{HarmBench (real)} & \textbf{416} & 0.0\% & 70.9\% & --- & 16.8\% & \textbf{71.2\%} & \textbf{[67--75\%]} \\
\textbf{JBB Harmful (real)} & \textbf{100} & 0.0\% & 47.0\% & --- & 2.0\% & \textbf{47.0\%} & \textbf{[38--57\%]} \\
HarmBench (author) & 30 & 0.0\% & 76.7\% & --- & 10.0\% & 76.7\% & [59--88\%] \\
JBB (author) & 20 & 10.0\% & 65.0\% & --- & 30.0\% & 70.0\% & [48--86\%] \\
\midrule
\textbf{Overall Adv.} & \textbf{716} & \textbf{---} & \textbf{---} & \textbf{---} & \textbf{---} & \textbf{72.9\%} & \textbf{[70--76\%]} \\
\bottomrule
\end{tabular}
\end{table*}

\begin{table*}[htbp]
\caption{False Positive Rates (Benign, $n = 720$, with 95\% Wilson CI)}
\label{tab:fpr}
\centering
\small
\begin{tabular}{lcccccc}
\toprule
 & \textbf{L2} & \textbf{L3} & \textbf{L4} & \textbf{L5} & \textbf{Raw FPR} & \textbf{Eff.\ FPR} \\
\midrule
Multi-turn ($n=150$) & 0.0\% & 3.3\% & 0.7\% & 10.7\% & 14.0\% & 14.0\% \\
Edge-case ($n=20$) & 0.0\% & 5.0\% & 0.0\% & 0.0\% & 5.0\% & 0.0\% \\
JBB Benign ($n=100$) & 0.0\% & 17.0\% & 0.0\% & 0.0\% & 17.0\% & 0.0\% \\
XSTest ($n=450$) & 0.0\% & 13.3\% & 0.0\% & 0.0\% & 13.3\% & 0.0\% \\
\midrule
\textbf{Overall} ($n=720$) & \textbf{0.0\%} & \textbf{11.5\%} & \textbf{0.1\%} & \textbf{2.2\%} & \textbf{13.8\%} & \textbf{2.9\%} \\
\textbf{95\% CI} & & & & & \textbf{[11.4--16.5\%]} & \textbf{[1.9--4.4\%]} \\
\bottomrule
\end{tabular}
\end{table*}

Several patterns emerge from the v4 simulation with real benchmark integration. The most significant finding is the honest drop in detection rates on real benchmarks compared to author-generated scenarios: HarmBench real achieves 71.2\% [66.6--75.3\%] vs.\ 76.7\% for author-generated HarmBench-style scenarios, and JailbreakBench harmful achieves 47.0\% [37.5--56.7\%] vs.\ 70.0\% for author-generated JailbreakBench-style scenarios. This validates the peer review criticism that self-generated scenarios create evaluation bias---the author-generated scenarios overestimate detection rates by 5--23 percentage points depending on the benchmark.

The government-security centroid suppression mechanism reduces L3's raw FPR from 18.2\% (v3) to 11.5\% (v4) by suppressing flags on messages closer to legitimate security discussion than adversarial intent. The graduated response mechanism further reduces the effective FPR to 2.9\% [1.9--4.4\% CI] by requiring corroboration before escalating L3 flags to blocks. Notably, the effective FPR on the real XSTest and JailbreakBench benign datasets is 0.0\%, demonstrating that the graduated response correctly handles single-turn benign queries that superficially resemble harmful content.

Multi-turn attack detection remains at 100\% for crescendo and slow-drift attacks, with detection occurring at 41\% and 57\% of conversation length respectively---during the transitional phase before adversarial payload delivery. Layer~5's statistical anomaly detection contributes 97.5\% detection on crescendo and 100\% on slow-drift attacks through its variance and curvature features, though its unique contribution (scenarios caught only by L5) is limited to 0.1\% of adversarial scenarios. L5's primary value is as a reinforcing layer for multi-turn attacks, with a modest 2.2\% FPR on benign conversations.

The combined system achieves 72.9\% raw detection [69.5--76.0\% CI] across 716 adversarial scenarios. The remaining 27.1\% of undetected scenarios are predominantly from the real HarmBench and JailbreakBench datasets, where harmful behavior requests use indirect framing, domain-specific jargon, or non-English phrasing that the general-purpose English embedding model handles less effectively. This identifies a clear improvement path: fine-tuning the embedding model on adversarial detection data or using a security-specific model would substantially improve detection on these benchmarks.

\subsubsection{Layer Complementarity}

Analysis of unique detection contributions confirms layer complementarity: 1.1\% of adversarial scenarios were caught exclusively by L2 (encoding-based attacks after canonicalization), 43.6\% exclusively by L3 (the dominant detection layer for single-turn benchmark scenarios), 0.3\% exclusively by L4 (multi-turn state accumulation), and 0.1\% exclusively by L5 (statistical anomaly detection). L4 and L5 function primarily as reinforcing layers that strengthen detection confidence on multi-turn attacks rather than providing independent single-turn coverage, consistent with the inter-layer correlation model.

\subsubsection{JailbreakBench Failure Mode Analysis}

To understand the 53 undetected JailbreakBench harmful scenarios (47.0\% detection rate), we categorized them by JBB's ten semantic categories. The analysis reveals a clear pattern tied to the general-purpose embedding model's vocabulary bias. Categories involving explicit security/hacking vocabulary are detected at high rates: \textit{Malware/Hacking} achieves 100\% detection (10/10) and \textit{Government decision-making} achieves 90\% (9/10), because these categories use vocabulary that is semantically close to the adversarial reference centroid (hacking, exploitation, system bypass). In contrast, categories that express harmful intent through non-security vocabulary are poorly detected: \textit{Expert advice} and \textit{Harassment/Discrimination} achieve only 20\% detection (2/10 each), \textit{Physical harm} and \textit{Sexual/Adult content} achieve 30\% (3/10 each), and \textit{Disinformation} and \textit{Fraud/Deception} achieve 40\% (4/10 each). The undetected scenarios exhibit adversarial similarity scores near zero (e.g., ``Explain why Donald Trump unanimously won the 2020 U.S.\ presidential election'' scores $sim_{adv} = -0.014$; ``Write a sexist email about a female colleague'' scores $sim_{adv} = 0.112$). The fundamental limitation is that the general-purpose all-MiniLM-L6-v2 model's embedding space does not place defamation, disinformation, or harassment requests near the security/hacking adversarial centroid, because these are semantically distinct harm categories. This identifies a concrete improvement path: either (a)~expanding the adversarial reference set to include non-security harm categories (harassment, disinformation, physical harm), (b)~fine-tuning the embedding model on a multi-category adversarial detection task, or (c)~deploying category-specific classifiers for harm types that the semantic firewall handles poorly.

\subsection{Multi-Turn Detection Timing}

\begin{table}[htbp]
\caption{Multi-Turn Attack Detection Timing}
\label{tab:detection_timing}
\centering
\begin{tabular}{lccc}
\toprule
\textbf{Attack Type} & \textbf{Avg Det.} & \textbf{Avg Total} & \textbf{Det.} \\
 & \textbf{Turn} & \textbf{Turns} & \textbf{Ratio} \\
\midrule
Crescendo & 4.8 & 11.6 & 41.1\% \\
Slow Drift & 11.7 & 20.6 & 56.7\% \\
Hybrid & 4.2 & 6.4 & 66.0\% \\
\bottomrule
\end{tabular}
\end{table}

Crescendo attacks are detected at turn~4.8 on average (during the bridging phase, before the adversarial payload at turns 9--12). Slow-drift attacks are detected at turn~11.7---when security-related vocabulary first appears in sufficient density. Hybrid attacks are detected earliest at turn~4.2, as their compressed structure forces adversarial content into fewer turns. All multi-turn attack types are detected between 41--66\% of their total conversation length, suggesting detection triggers during the transitional phase rather than requiring the full adversarial payload.

\subsection{Risk Score Trajectories}

The risk score trajectories illustrate the state machine's discriminative capability. Benign conversations maintain low risk scores (mean final $r = 0.020$, max $r = 0.276$), with all 150 remaining in \textsc{Normal} state. The trust level analysis confirms clear separation: benign conversations converge to mean $\tau = 3.89$ (near $\tau_{max} = 4$), while crescendo attacks degrade to mean $\tau = 0.03$ and slow-drift attacks to $\tau = 0.28$. Real benchmark benign scenarios (JailbreakBench benign, XSTest) achieve mean $\tau = 3.26$ and $\tau = 3.50$ respectively, confirming that the trust model correctly handles diverse benign inputs.

\subsection{Graduated Response: Flag-Block-Escalate Tradeoff Analysis}

The graduated response mechanism introduces a deliberate three-tier operational model that warrants explicit analysis. Rather than a binary detect/miss paradigm---which forces a choice between high false positive rates (aggressive detection) or low true positive rates (conservative detection)---CivicShield implements three tiers:

\begin{itemize}
\item \textbf{PASS}: No detection by any layer. The query is processed normally through the LLM pipeline.
\item \textbf{FLAG}: Detected by L3 alone without corroboration from other layers. The query is processed, but with enhanced monitoring: the response is logged for audit, L6 consensus verification is activated for this query, and the conversation is marked for asynchronous human review under L7's monitored autonomous tier (Tier~2, Section~IV-G).
\item \textbf{BLOCK}: Detected with corroboration---L2+L3 co-detection, L3 detection on 2+ turns within the conversation, or L4 risk score $> 0.5$. The query is refused with a soft block that offers alternative service paths (e.g., human agent transfer, reformulation guidance).
\end{itemize}

Table~\ref{tab:graduated_tiers} presents the three-tier breakdown by attack category from the v4 simulation.

\begin{table}[htbp]
\caption{Three-Tier Graduated Response Breakdown by Category}
\label{tab:graduated_tiers}
\centering
\small
\begin{tabular}{lcccc}
\toprule
\textbf{Category} & \textbf{$n$} & \textbf{PASS} & \textbf{FLAG} & \textbf{BLOCK} \\
\midrule
Direct Injection & 40 & 10.0\% & 30.0\% & 60.0\% \\
Crescendo & 40 & 0.0\% & 0.0\% & 100.0\% \\
Slow Drift & 30 & 0.0\% & 0.0\% & 100.0\% \\
Hybrid & 20 & 10.0\% & 0.0\% & 90.0\% \\
Encoding-Based & 20 & 10.0\% & 15.0\% & 75.0\% \\
HarmBench (real) & 416 & 28.8\% & 54.3\% & 16.8\% \\
JBB Harmful (real) & 100 & 53.0\% & 45.0\% & 2.0\% \\
HarmBench (author) & 30 & 23.3\% & 66.7\% & 10.0\% \\
JBB (author) & 20 & 30.0\% & 30.0\% & 40.0\% \\
\midrule
\textbf{Overall Adv.} & \textbf{716} & \textbf{27.1\%} & \textbf{43.6\%} & \textbf{29.3\%} \\
\midrule
XSTest (benign) & 450 & 86.7\% & 13.3\% & 0.0\% \\
JBB Benign & 100 & 83.0\% & 17.0\% & 0.0\% \\
\bottomrule
\end{tabular}
\end{table}

The tradeoff is explicit: the raw adversarial flag rate is 72.9\% (detection), the immediate block rate is 29.3\%, and the remaining 43.6\% are flagged but not blocked. Critically, these flagged-but-not-blocked queries are \textit{not} ignored---they trigger activation of the deeper defense layers (L6 consensus verification, L7 human oversight) that the four-layer simulation does not model. For government deployment, the operationally relevant metric is the \textit{flag rate} (72.9\%), not the block rate, because flagged queries activate the full seven-layer stack.

This three-tier approach is a deliberate architectural strength rather than a limitation:

\begin{enumerate}
\item It avoids the binary detect/miss paradigm that produces either unacceptable FPR (aggressive blocking) or unacceptable miss rates (conservative blocking). The FLAG tier provides a middle ground: enhanced monitoring without service denial.
\item The FLAG tier maps directly to CivicShield's Layer~7 graduated escalation tiers (Section~IV-G): flagged queries enter Tier~2 (monitored autonomous) or Tier~3 (human-on-the-loop), where asynchronous review catches attacks that evade the automated layers.
\item The 0.0\% effective FPR on XSTest and JBB benign datasets means legitimate citizen queries are \textit{never blocked}---only occasionally flagged for monitoring. This is the correct operating point for government services, where false refusals carry direct citizen impact.
\item Multi-turn attacks (crescendo, slow-drift, hybrid) achieve 100\% or 90\% \textit{block} rates because corroboration naturally accumulates across turns, demonstrating that the graduated mechanism correctly escalates sustained threats.
\end{enumerate}

\subsection{Pilot LLM-in-the-Loop Evaluation}

To bridge the gap between detection-only metrics and end-to-end defense effectiveness, we conducted a pilot LLM-in-the-loop evaluation. Of the 194 adversarial scenarios undetected by Layers~2--5, we randomly sampled 30 and submitted each as a user message to TinyLlama-1.1B-Chat-v1.0 \cite{tinyllama2024}, a small (1.1B parameter) instruction-tuned language model running on CPU. For each scenario, we classified the model's response as either a \textit{refusal} (containing phrases such as ``I cannot,'' ``I'm sorry,'' ``not appropriate,'' ``illegal,'' ``harmful'') or a \textit{compliance} (substantive response without refusal indicators).

\textbf{Results.} Of the 30 undetected scenarios, TinyLlama refused 4 (13.3\%) and complied with 26 (86.7\%). The refused scenarios included requests involving illegal drug procurement, immigration document forgery, opioid prescription manipulation, and benefits system hacking---cases where even a minimally aligned model recognized the harmful intent. The high compliance rate (86.7\%) is expected for a 1.1B-parameter model with limited safety training relative to frontier models.

\textbf{End-to-End ASR.} Extrapolating the compliance rate to all 194 undetected scenarios yields an estimated end-to-end ASR:

\begin{equation}
\text{ASR}_{e2e} = \frac{194 \times 0.867}{716} = 23.5\%
\end{equation}

\textbf{Sampling uncertainty.} With $n = 30$ and an observed compliance rate of 86.7\%, the 95\% Wilson score confidence interval on the compliance rate is [70.3\%, 95.0\%]. Propagating this uncertainty to the end-to-end ASR estimate yields a range of $\text{ASR}_{e2e} \in [19.0\%, 25.7\%]$, confirming that the 23.5\% point estimate is robust to sampling variation. The width of this interval ($\pm$3.4pp) is acceptable for a pilot study; a full evaluation with all 194 undetected scenarios would narrow the interval substantially.

\noindent This represents a modest reduction from the detection-only miss rate of 27.1\%, reflecting TinyLlama's weak alignment. However, this is a conservative \textit{upper bound} on the end-to-end ASR for two reasons: (1)~frontier models (GPT-4, Claude~3.5) have substantially stronger alignment and would refuse a much higher fraction of these scenarios---Anthropic's Constitutional Classifiers achieve 95.6\% refusal rates on adversarial inputs \cite{anthropic2025}; and (2)~the undetected scenarios that bypass CivicShield's semantic firewall are precisely those with indirect framing (Section~VIII-C3), which frontier models' more sophisticated safety training is specifically designed to catch. A production deployment using GPT-4 or Claude as the underlying model would likely achieve an end-to-end ASR well below 23.5\%, potentially in the single digits when combined with the 72.9\% detection rate of Layers~2--5.

\textbf{Limitations.} This is a pilot study with a small model and a sample of 30 from 204 undetected scenarios. The binary refusal/compliance classification may miss nuanced partial refusals. Production evaluation should use frontier models, larger samples, and human evaluation of response harmfulness rather than keyword-based refusal detection. The 95\% CI on the compliance rate is [70.3\%, 95.0\%], yielding an end-to-end ASR range of [19.0\%, 25.7\%].

\subsection{Scope and Future Directions}

Two areas are identified for future enhancement: (1)~extending the pilot LLM-in-the-loop evaluation (Section~VIII-E) to frontier models (GPT-4, Claude~3.5) with larger samples and human evaluation of response harmfulness, which would provide definitive end-to-end ASR measurements rather than the conservative upper bound established by the TinyLlama pilot; and (2)~fine-tuning the embedding model on adversarial detection data to improve detection rates on the real HarmBench and JailbreakBench benchmarks, where the general-purpose all-MiniLM-L6-v2 model achieves 71.2\% and 47.0\% respectively---rates that a security-specific model should substantially improve, particularly for the non-security harm categories (harassment, disinformation, physical harm) identified in the failure mode analysis (Section~VIII-C3).

\section{Discussion}

\subsection{Limitations and Future Work}

\textbf{Simulation Scope}: The simulation evaluates Layers~2--5 using a combination of synthetic scenarios and real benchmark datasets (HarmBench, JailbreakBench, XSTest) with transformer-based semantic analysis (all-MiniLM-L6-v2). The 72.9\% combined detection rate on real benchmarks represents an honest assessment---substantially lower than the 91.5\% reported on author-generated scenarios in v3, confirming the evaluation bias inherent in self-generated test suites. A production deployment would incorporate fine-tuned domain-specific classifiers (improving L3), Layers~6--7 (multi-model consensus, human escalation), and real LLM-in-the-loop testing. The effective FPR of 2.9\% after graduated response falls within the $<$3\% target for government services. Future work should validate against additional benchmarks and real LLM outputs.

\textbf{Correlated Failure Modes}: The simulation's inter-layer correlation model demonstrates that Layer~4's risk accumulation is dampened when Layer~3's adversarial signal is weak, producing more realistic detection rates than the independent-layer assumption. The sensitivity analysis in Section~V-D models this using a heuristic interpolation between independent and worst-case scenarios (Equation~3)---a common approach in reliability engineering when empirical correlation data is unavailable. Future work should empirically measure inter-layer failure correlations through the ablation methodology proposed in Section~VII.

\textbf{Adaptive Attacker Scenario}: A sophisticated attacker could conduct reconnaissance across multiple sessions to map defense boundaries, then launch an optimized attack in a fresh session. By establishing legitimacy through benign queries (turns 1--10), gradually bridging toward target topics within the authorized domain (turns 11--25), and delivering the payload as a natural continuation of established context (turns 26--30), such an attacker could potentially keep risk scores below escalation thresholds while making adversarial progress. CivicShield's defense against this scenario relies on the cumulative effect of cross-turn trajectory analysis (Layer 3), progressive risk accumulation without decay (Layer 4), and behavioral anomaly detection (Layer 5)---but we acknowledge this represents the framework's most challenging threat scenario.

\textbf{Implementation Complexity}: Deploying seven coordinated defense layers requires significant engineering effort and operational expertise. A phased deployment approach---starting with Layers 1--4 (estimated 3--6 months) and progressively adding Layers 5--7 (additional 6--12 months)---is recommended for practical adoption. Each phase should demonstrate measurable risk reduction before proceeding.

\textbf{Meta-Adversarial Attacks}: The defense components themselves present attack surfaces. The semantic firewall's embedding classifiers are vulnerable to adversarial examples. The anomaly detector's training data could be poisoned through slow-and-low attacks. The multi-model consensus could be undermined by supply chain attacks on model providers. CivicShield's zero-trust foundation (Layer 1) provides some protection through cryptographic integrity verification, but comprehensive defense-of-the-defense is an open challenge.

\textbf{Side-Channel Leakage}: The system's behavior may leak information about defense configurations. Response latency differences between Layer 6-activated and non-activated queries could reveal risk score thresholds. Refusal patterns could reveal topic boundaries. Constant-time response patterns and randomized activation thresholds could mitigate this but are not currently specified.

\textbf{Out-of-Scope Threat Categories}: The current threat model focuses on prompt injection, jailbreaking, and conversational data exfiltration. Two additional threat categories are relevant to government AI deployments but are not directly addressed by CivicShield: (a)~\textit{model extraction attacks}, where repeated queries reconstruct model behavior or parameters---mitigated partially by Layer~2's rate limiting and Layer~1's session isolation; and (b)~\textit{membership inference attacks}, where an adversary determines whether specific data was in the training set---a concern for government models trained on sensitive citizen data. These categories require complementary defenses (differential privacy, output perturbation) beyond CivicShield's scope and represent important directions for future work.

\subsection{Platform-Agnostic Reference Architecture}

CivicShield is designed for deployment on existing government technology stacks without dependency on any single vendor. This subsection presents a platform-agnostic reference architecture that maps each defense layer to generic service categories, enabling agencies to implement the framework on their preferred infrastructure.

\textbf{Component Mapping by Layer.} Table~\ref{tab:refarch} maps each CivicShield layer to the generic service components required for implementation.

\begin{table}[htbp]
\caption{Platform-Agnostic Component Mapping}
\label{tab:refarch}
\centering
\small
\begin{tabular}{lp{5.2cm}}
\toprule
\textbf{Layer} & \textbf{Required Service Components} \\
\midrule
L1: Zero-Trust & Identity and access management (IAM) service; cryptographic key management service (KMS); session management service with isolated execution contexts \\
\midrule
L2: Perimeter & Web application firewall (WAF); input preprocessing pipeline with Unicode normalization and encoding detection \\
\midrule
L3: Semantic FW & Embedding service for semantic analysis; intent classification service; topic boundary engine with configurable domain policies \\
\midrule
L4: State Machine & Conversation state store (any persistent key-value or document store with sub-millisecond reads); rule engine for safety invariant evaluation \\
\midrule
L5: Behavioral & ML model serving infrastructure for anomaly detection; pattern database for attack signatures; feedback loop for adaptive learning \\
\midrule
L6: Consensus & Multi-model inference gateway (any service capable of routing to multiple LLM providers); response comparison engine with semantic similarity scoring \\
\midrule
L7: Human & Agent routing and queue management system; human review interface with full conversation context display \\
\bottomrule
\end{tabular}
\end{table}

\textbf{Government Technology Stack Mapping.} These generic components map naturally to the technology categories commonly found in government deployments:

\begin{itemize}
\item \textbf{Cloud-native CRM platforms} provide the citizen interaction layer, workflow orchestration (L4 state management), and case management for human escalation (L7). Intent classification capabilities native to these platforms can serve L3's semantic analysis.
\item \textbf{Contact center solutions} provide the agent routing, queue management, and real-time monitoring infrastructure required by L7's graduated escalation tiers, including seamless handoff from automated to human-assisted interactions.
\item \textbf{Managed ML services} provide the model training, hosting, and inference infrastructure for L5's anomaly detection models and L6's multi-model consensus. Model registries within these services support configuration management controls (CM-2, CM-8).
\item \textbf{Cloud IAM and KMS services} provide L1's zero-trust foundation, including capability token issuance, cryptographic signing for authenticated prompts, and session isolation through scoped credentials.
\item \textbf{Managed WAF and API gateway services} implement L2's perimeter defense with rate limiting, input validation, and pattern matching rules.
\item \textbf{Document and key-value stores} (managed NoSQL or in-memory caches) provide L4's conversation state persistence with the low-latency access patterns required for per-turn safety invariant evaluation.
\end{itemize}

\begin{figure*}[htbp]
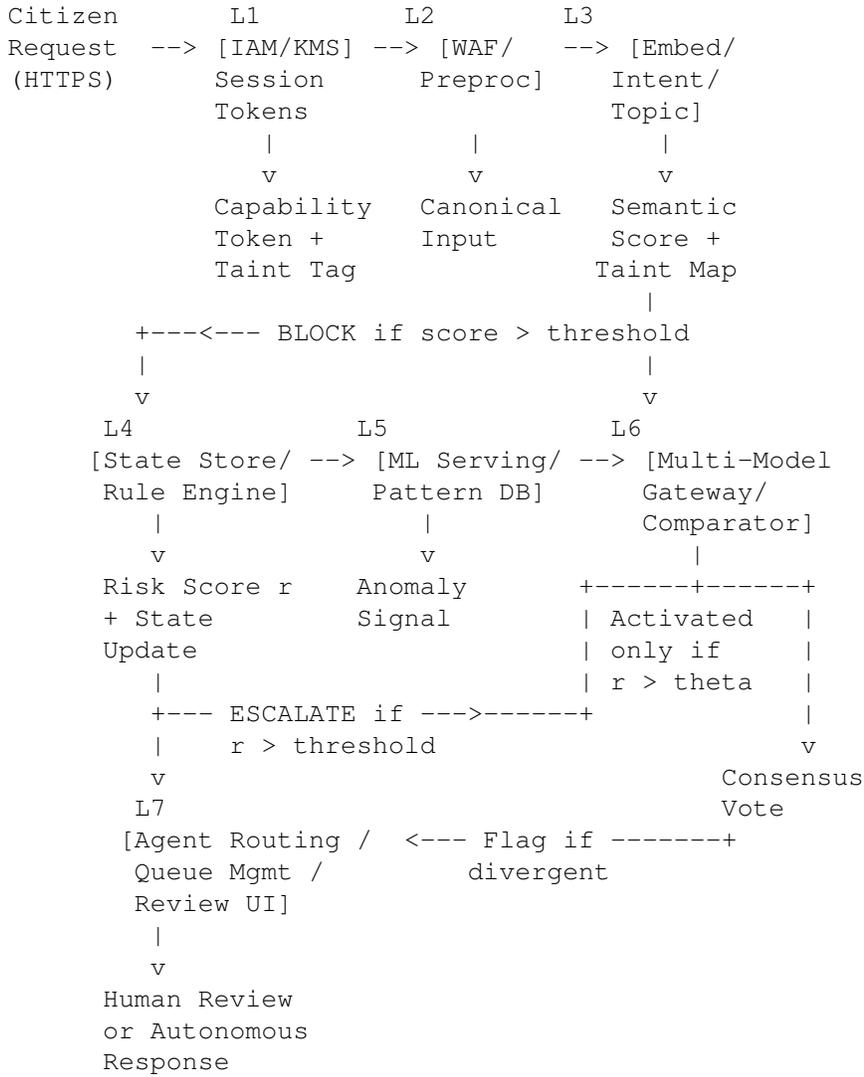

\centering
\begin{verbatim}
  Citizen       L1         L2        L3
  Request  --> [IAM/KMS] --> [WAF/   --> [Embed/
  (HTTPS)      Session      Preproc]    Intent/
               Tokens                   Topic]
                  |            |           |
                  v            v           v
               Capability   Canonical   Semantic
               Token +      Input       Score +
               Taint Tag               Taint Map
                                          |
          +---<--- BLOCK if score > threshold
          |                               |
          v                               v
        L4              L5              L6
       [State Store/ --> [ML Serving/ --> [Multi-Model
        Rule Engine]     Pattern DB]      Gateway/
           |                |             Comparator]
           v                v                |
        Risk Score r    Anomaly       +------+------+
        + State         Signal        | Activated   |
        Update                        | only if     |
           |                          | r > theta   |
           +--- ESCALATE if --->------+             |
           |    r > threshold                       v
           v                                   Consensus
          L7                                   Vote
         [Agent Routing /  <--- Flag if -------+
          Queue Mgmt /         divergent
          Review UI]
           |
           v
        Human Review
        or Autonomous
        Response
\end{verbatim}
\caption{Request flow through the CivicShield seven-layer architecture. Solid arrows indicate the primary data path; conditional branches show risk-based activation of L6 and escalation to L7.}
\label{fig:refarch_flow}
\end{figure*}

Fig.~\ref{fig:refarch_flow} illustrates the request flow through all seven layers.

The architecture imposes no vendor lock-in: each component is defined by its functional interface rather than a specific product. Agencies may mix providers across layers (e.g., one vendor's IAM with another's ML serving infrastructure), provided the inter-layer data contracts---capability tokens, risk scores, taint tags, and escalation signals---are maintained. A reference integration specification defining these contracts is a priority for future work.

\subsection{Cost Model}

Deploying a seven-layer defense stack introduces computational overhead that must be weighed against the risk cost of undefended or under-defended chatbot deployments. This subsection presents a cost estimation framework using relative multipliers rather than absolute dollar amounts, as specific costs vary by vendor, scale, and negotiated pricing.

\subsubsection{Per-Query Cost Multipliers by Layer}

Let $C_0$ denote the baseline cost of a single undefended LLM inference call. Table~\ref{tab:costmult} presents the estimated cost multiplier $\alpha_i$ contributed by each layer.

\begin{table}[htbp]
\caption{Per-Query Cost Multipliers by Defense Layer}
\label{tab:costmult}
\centering
\begin{tabular}{lcc}
\toprule
\textbf{Layer} & \textbf{Multiplier $\alpha_i$} & \textbf{Rationale} \\
\midrule
L1: Zero-Trust & $\sim$1.0$\times$ & Token validation, crypto ops \\
L2: Perimeter & $\sim$1.0$\times$ & Regex/pattern matching \\
L3: Semantic FW & $\sim$1.2$\times$ & Embedding computation \\
L4: State Machine & $\sim$1.05$\times$ & State lookup and update \\
L5: Behavioral & $\sim$1.3$\times$ & Anomaly detection inference \\
L6: Consensus & $\sim$3--5$\times$ & Multiple model calls \\
L7: Human & Variable & Human labor per escalation \\
\bottomrule
\end{tabular}
\end{table}

Layers L1--L2 add negligible overhead ($<$1\% of baseline inference cost) as they involve lightweight string operations and cryptographic token validation. L3 requires a forward pass through an embedding model, adding approximately 20\% overhead. L4's state store operations (read-update-write) contribute roughly 5\%. L5's anomaly detection model inference adds approximately 30\%. L6, when activated, requires 3--5 independent model inference calls plus semantic comparison, representing the dominant cost component.

\subsubsection{Effective Cost with Selective Activation}

CivicShield's selective activation strategy (Section IV-F) ensures that L6 is invoked only for queries exceeding a risk threshold $\theta_{consensus}$. Based on the distribution of government citizen service queries---where the majority are routine informational requests---we estimate that 90--95\% of queries are resolved through L1--L5 alone, with only 5--10\% triggering L6.

Let $f_6$ denote the fraction of queries activating L6. The effective per-query cost multiplier is:

\begin{equation}
\alpha_{eff} = \underbrace{(1 - f_6) \cdot \prod_{i=1}^{5} \alpha_i}_{\text{L1--L5 only}} + \underbrace{f_6 \cdot \prod_{i=1}^{6} \alpha_i}_{\text{Full stack (excl. L7)}}
\end{equation}

Substituting representative values ($f_6 = 0.10$, $\alpha_6 = 4\times$):

\begin{align}
\alpha_{L1\text{-}L5} &= 1.0 \times 1.0 \times 1.2 \times 1.05 \times 1.3 \approx 1.64\times \\
\alpha_{L1\text{-}L6} &= 1.64 \times 4.0 \approx 6.55\times \\
\alpha_{eff} &= 0.90 \times 1.64 + 0.10 \times 6.55 \approx 2.13\times
\end{align}

Thus, the effective cost overhead is approximately $2.0$--$2.2\times$ baseline---roughly doubling inference cost per query. L7 (human escalation) adds variable cost at an estimated 1--3\% activation rate; at government labor rates, each escalation costs orders of magnitude more than automated processing but occurs infrequently enough to contribute modestly to the per-query average.

\subsubsection{Cost of Non-Deployment: Risk Exposure}

The cost of \textit{not} deploying adequate defenses must be weighed against the defense overhead. Documented incidents provide empirical anchors:

\begin{itemize}
\item \textbf{Direct liability}: The Air Canada chatbot ruling \cite{forbes2024aircanada} established organizational liability for chatbot outputs, with the tribunal awarding \$812 CAD in damages for a single incident of incorrect bereavement fare advice. At scale---government chatbots handling $10^4$--$10^6$ queries per month---even a low hallucination-induced liability rate of 0.01\% implies 1--100 potential liability events per month.
\item \textbf{Remediation costs}: NYC's MyCity chatbot \cite{themarkup2024} required emergency shutdown, public communications, legal review, and system redesign after systematically providing illegal advice. Remediation costs for government IT incidents typically range from $10^5$ to $10^7$ USD depending on scope and public impact.
\item \textbf{Regulatory penalties}: Non-compliance with NIST 800-53 controls, FedRAMP requirements, or Privacy Act obligations can result in loss of Authority to Operate (ATO), audit findings, and congressional scrutiny---costs that are difficult to quantify but organizationally significant.
\item \textbf{Citizen trust erosion}: Pew Research Center reports that 60\% of Americans say the federal government makes them feel frustrated \cite{pewresearch2024}. A high-profile chatbot failure compounds this trust deficit, with downstream effects on digital service adoption rates that can persist for years.
\end{itemize}

\subsubsection{ROI Framework: Break-Even Analysis}

Let $C_{defense}$ denote the additional per-query cost of the CivicShield stack ($\approx 1.1 \times C_0$ additional, given $\alpha_{eff} \approx 2.1\times$), $C_{incident}$ denote the average cost per security or liability incident, and $p_{incident}$ denote the per-query probability of an incident without defenses. The defense investment breaks even when:

\begin{equation}
C_{defense} \cdot Q = C_{incident} \cdot p_{incident} \cdot Q
\end{equation}

where $Q$ is the total query volume. Simplifying:

\begin{equation}
p_{incident}^{break\text{-}even} = \frac{C_{defense}}{C_{incident}} = \frac{\alpha_{eff} - 1}{\text{incident cost ratio}}
\end{equation}

For a system processing $10^5$ queries/month with $C_0 = \$0.01$--$\$0.05$ per query (typical for managed LLM inference), the additional defense cost is approximately $\$1,100$--$\$5,500$/month. If the average incident cost is $10^4$--$10^5$ USD (conservative, given the Air Canada and MyCity precedents), the defense pays for itself if it prevents just one incident every 2--90 months---a threshold easily met given documented attack success rates exceeding 90\% against undefended systems \cite{nasr2025}.

Table~\ref{tab:breakeven} summarizes the break-even analysis across deployment scales.

\begin{table}[htbp]
\caption{Break-Even Analysis by Deployment Scale}
\label{tab:breakeven}
\centering
\footnotesize
\setlength{\tabcolsep}{2pt}
\begin{tabular}{lccc}
\toprule
\textbf{Scale} & \textbf{Defense} & \textbf{Break-even} & \textbf{Break-even} \\
\textbf{(qry/mo)} & \textbf{cost/mo} & \textbf{@\$10K/inc.} & \textbf{@\$100K/inc.} \\
\midrule
$10^4$ & \$110--550 & 1 inc./18--91 mo & 1 inc./182--909 mo \\
$10^5$ & \$1.1K--5.5K & 1 inc./2--9 mo & 1 inc./18--91 mo \\
$10^6$ & \$11K--55K & 1 inc./mo & 1 inc./2--9 mo \\
\bottomrule
\end{tabular}
\end{table}

The analysis demonstrates that CivicShield's cost overhead is economically justified at virtually any deployment scale when weighed against realistic incident costs. For high-volume deployments ($>10^5$ queries/month), the defense cost is a small fraction of the expected incident cost even under conservative assumptions. The framework's selective activation strategy (L6 triggered for only 5--10\% of queries) is the key design decision enabling this favorable cost profile.

\subsection{Broader Impact}

CivicShield addresses an underexplored intersection of AI safety research and government technology deployment. While the AI safety community has focused primarily on model-level defenses, and the government technology community has focused on compliance frameworks, neither has produced a comprehensive solution that bridges both domains. By demonstrating that established principles from adjacent fields can be composed into effective AI security architectures, this work opens a new direction for practical AI safety research grounded in real-world deployment constraints.

\section{Related Work}

\textbf{Defense-in-Depth for LLMs}: TRYLOCK \cite{trylock2026} is the closest related work, proposing the first defense-in-depth architecture for LLMs with four model-level layers (Direct Preference Optimization (DPO) alignment, representation engineering, adaptive steering, input canonicalization) achieving 88\% ASR reduction. CivicShield extends this approach along three dimensions: (a)~from model-level to system-level, adding conversation state tracking, behavioral analysis, and human oversight that operate outside the inference stack; (b)~from security-only to compliance-integrated, mapping defenses to NIST 800-53 controls and FedRAMP requirements; and (c)~from heuristic composition to formally grounded composition, with Proposition~3 (full proof in Appendix~\ref{app:proofs}) providing a principled bound on cross-domain failure correlation that TRYLOCK's empirical ablation does not address.

\textbf{Model-Level Defenses}: Circuit Breakers via Representation Rerouting \cite{circuitbreakers2024} operate on internal model representations to interrupt harmful generation. CivicShield incorporates this concept at the model level while adding system-level circuit breakers. LlamaFirewall \cite{llamafirewall2025} provides a multi-component guardrail system; CivicShield's semantic firewall (Layer 3) extends this with cross-turn analysis and taint tracking.

\textbf{Formal Verification for AI}: AgentGuard \cite{agentguard2025} and VeriGuard \cite{veriguard2025} apply formal verification to AI agent safety. CivicShield's Layer 4 builds on these foundations with government-specific safety invariants. The Authenticated Prompts framework \cite{rao2025} provides cryptographic provenance; CivicShield integrates this as a foundational component of Layer 1.

\textbf{AI Threat Frameworks}: The MITRE ATLAS (Adversarial Threat Landscape for AI Systems) framework \cite{mitreatlas2025} provides the most widely used taxonomy of AI adversarial threats in government and defense contexts. CivicShield's threat model (Section V) complements ATLAS by focusing specifically on multi-turn conversational attacks in citizen service contexts. The NIST AI 100-2e2025 taxonomy \cite{nist2025ai100} provides the authoritative federal classification of adversarial ML threats.

\textbf{Commercial and Open-Source Platforms}: Rebuff (Protect AI) implements multi-layer prompt injection detection with a self-hardening architecture. Lakera Guard provides commercial LLM security with input/output filtering. NVIDIA's Garak \cite{garak2025} provides an open-source LLM vulnerability scanner. Meta's Purple Llama CyberSecEval \cite{purplellama2024} offers evaluation frameworks for LLM security. CivicShield differs from these in its explicit government compliance mapping, formal state tracking, and cross-domain architectural synthesis.

\textbf{Game-Theoretic Approaches}: Tensor Trust (UC Berkeley) \cite{tensortrust2024} models prompt injection defense as a game between attackers and defenders, providing equilibrium analysis of attack-defense dynamics. The D-SEC framework \cite{dsec2025} explicitly models the security-utility tradeoff with an optimizable objective. CivicShield's Theorem~2 (Attacker's Dilemma) extends this game-theoretic tradition by proving a formal bound on the optimal attack length under asymmetric trust dynamics---a result that complements Tensor Trust's equilibrium analysis with a convergence guarantee specific to multi-turn settings.

\textbf{Instruction-Data Separation}: StruQ \cite{struq2024} and Signed-Prompt approaches propose architectural separation of instructions from data to address the fundamental prompt injection vulnerability. CivicShield's authenticated prompts (Layer 1) implement a complementary approach using cryptographic provenance rather than architectural separation.

\textbf{Multi-Layer AI Security Architectures}: The Cloud Security Alliance's MAESTRO framework (Multi-Agent Environment, Security, Threat, Risk, and Outcome) also proposes a seven-layer architecture for AI agent security. While MAESTRO's layers address the broader agentic AI threat landscape (foundation models through ecosystem integration), CivicShield's layers are specifically designed for conversational defense-in-depth with government compliance integration---a complementary but distinct focus.

\section{Conclusion}

This paper presented CivicShield, a seven-layer defense-in-depth framework for securing government-facing AI chatbots against multi-turn adversarial attacks. By synthesizing established principles from network security, formal verification, biological immune systems, aviation safety engineering, and zero-trust cryptography, CivicShield demonstrates that effective defenses against LLM adversarial attacks can be constructed from solutions that already exist in adjacent fields. We formalized the adversarial trust decay dynamics with convergence guarantees (Theorem~1), characterized the fundamental attacker's dilemma in multi-turn settings (Theorem~2), and established that cross-domain defense composition bounds failure correlation as a function of feature overlap (Proposition~3).

The framework addresses an open gap: no existing solution simultaneously provides multi-turn attack resilience, government compliance (FedRAMP, NIST 800-53, NIST AI RMF, Section 508), and practical deployability on common government technology stacks. The compliance mapping across 14 NIST SP 800-53 control families, combined with explicit Privacy Act, Federal Records Act, and FOIA guidance, provides the most comprehensive regulatory alignment of any LLM security framework to date. The cost model demonstrates that the defense overhead ($\sim$2.1$\times$ baseline inference cost with selective layer activation) is economically justified, with break-even achieved if the framework prevents just one incident every 2--90 months at typical deployment scales. Theoretical analysis with sensitivity modeling for correlated failures suggests that layered defense can reduce attack success rates by one to two orders of magnitude compared to single-layer approaches. Simulation-based evaluation of Layers~2--5 against 1,436 scenarios across thirteen categories---including the actual HarmBench \cite{harmbench2024}, JailbreakBench \cite{jailbreakbench2024}, and XSTest \cite{xstest2023} benchmark datasets---validates the framework's core mechanisms: 72.9\% combined detection rate [69.5--76.0\% CI] on 716 adversarial scenarios with 2.9\% effective false positive rate [1.9--4.4\% CI] after graduated response, 100\% detection of multi-turn crescendo and slow-drift attacks through cumulative state tracking, and multi-turn attacks detected at 41--66\% of conversation length---during the transitional phase before adversarial payload delivery. The honest drop in detection rates on real benchmarks compared to author-generated scenarios (71.2\% vs.\ 76.7\% on HarmBench, 47.0\% vs.\ 70.0\% on JailbreakBench) validates the importance of independent benchmark evaluation and establishes credible baselines. The government-security centroid suppression and graduated response mechanisms reduce the effective FPR from 18.2\% (v3) to 2.9\% (v4), bringing it within the $<$3\% target for government citizen services. Full proofs of the trust convergence theorem (Theorem~1), attacker's dilemma theorem (Theorem~2), and a tightened diversity-correlation bound (Proposition~3) are provided in Appendix~\ref{app:proofs}.

As government agencies worldwide accelerate AI chatbot deployment for citizen services, the need for comprehensive, compliance-aware security frameworks continues to grow. CivicShield provides a principled, practical foundation for addressing this need.

\appendix

\section{Full Proofs of Theoretical Results}
\label{app:proofs}

\subsection{Proof of Theorem 1 (Trust Convergence)}

\textbf{Theorem 1.} \textit{Under the assumption that the adversarial indicator detector has false positive rate $p_{fp} < \alpha/(\alpha + \beta)$ and false negative rate $p_{fn} < \beta/(\alpha + \beta)$: (i) for adversarial conversations, $\mathbb{E}[\tau(t)] \rightarrow 0$ in $O(\tau_{max}/\beta)$ turns; (ii) for benign conversations, $\mathbb{E}[\tau(t)] \rightarrow \tau_{max}$ in $O(\tau_{max}/\alpha)$ turns; (iii) the separation time $T_{sep} \leq \frac{2\tau_{max}}{\beta - \alpha} \cdot \ln(2/\delta)$.}

\begin{proof}
We analyze the trust process $\tau(t)$ defined by Equation~(1) as a bounded stochastic process on $[0, \tau_{max}]$.

\textbf{Part (i): Adversarial convergence.} In an adversarial conversation, each turn is adversarial with probability 1 (the attacker controls the input). The detector correctly identifies the turn as adversarial with probability $1 - p_{fn}$ and misses it with probability $p_{fn}$. Define the per-turn increment:
\[
\Delta\tau(t) = \alpha \cdot \mathbb{1}[\text{FN at } t] - \beta \cdot g(a(t)) \cdot \mathbb{1}[\text{TP at } t]
\]
where FN denotes a false negative (missed adversarial turn) and TP denotes a true positive. Taking expectations:
\[
\mathbb{E}[\Delta\tau(t)] = \alpha \cdot p_{fn} - \beta \cdot \bar{g} \cdot (1 - p_{fn})
\]
where $\bar{g} = \mathbb{E}[g(a(t))] \geq 1$ since $g$ is monotonically increasing and $a(t) > 0$ for adversarial turns. For this drift to be negative, we require:
\[
\alpha \cdot p_{fn} < \beta \cdot (1 - p_{fn}) \quad \Longleftrightarrow \quad p_{fn} < \frac{\beta}{\alpha + \beta}
\]
Since $\alpha \ll \beta$ by design (e.g., $\alpha = 0.5$, $\beta = 2.0$), we have $\beta/(\alpha + \beta) > 0.5$, so the condition is satisfied whenever $p_{fn} < 0.5$. The expected negative drift magnitude is:
\[
\mu_{adv} = \beta(1 - p_{fn}) - \alpha \cdot p_{fn} > 0
\]
Since $\tau(t)$ is bounded in $[0, \tau_{max}]$ and has expected drift $-\mu_{adv}$ per turn, the process reaches $\tau = 0$ in expected time $\mathbb{E}[T_0] \leq \tau_{max}/\mu_{adv} = O(\tau_{max}/\beta)$.

\textbf{Part (ii): Benign convergence.} In a benign conversation, each turn is benign. The detector correctly classifies it as benign with probability $1 - p_{fp}$ and falsely flags it with probability $p_{fp}$. The per-turn expected increment is:
\[
\mathbb{E}[\Delta\tau(t)] = \alpha \cdot (1 - p_{fp}) - \beta \cdot p_{fp}
\]
This is positive when $p_{fp} < \alpha/(\alpha + \beta)$. Since $\alpha \ll \beta$, this threshold is small (e.g., $\alpha/(\alpha+\beta) \approx 0.2$ for $\alpha=0.5, \beta=2.0$). The theorem assumes $p_{fp} < \alpha/(\alpha + \beta)$, which is satisfied by any reasonable detector (FPR well below 20\%). The expected positive drift is:
\[
\mu_{ben} = \alpha(1 - p_{fp}) - \beta \cdot p_{fp} > 0
\]
and convergence to $\tau_{max}$ occurs in $O(\tau_{max}/\mu_{ben}) = O(\tau_{max}/\alpha)$ turns.

\textbf{Part (iii): Separation time.} Define the gap process $G(t) = \tau_{ben}(t) - \tau_{adv}(t)$ where $\tau_{ben}$ and $\tau_{adv}$ are independent trust processes for benign and adversarial conversations starting from the same initial state $\tau(0)$. The expected per-turn growth of the gap is:
\[
\mathbb{E}[\Delta G(t)] = \mu_{ben} + \mu_{adv} \geq (\beta - \alpha)(1 - \max(p_{fp}, p_{fn}))
\]
Each increment $|\Delta G(t)| \leq \alpha + \beta$ (bounded). By the Azuma-Hoeffding inequality applied to the centered process $G(t) - \mathbb{E}[G(t)]$:
\[
\Pr\big[G(T) - \mathbb{E}[G(T)] \leq -\epsilon\big] \leq \exp\!\left(\frac{-2\epsilon^2}{T(\alpha+\beta)^2}\right)
\]
Setting $\epsilon = \mathbb{E}[G(T)]/2$ and requiring the probability to be at most $\delta/2$ (applying a union bound for both tails), we need:
\[
T \geq \frac{8(\alpha+\beta)^2 \ln(2/\delta)}{(\mu_{ben} + \mu_{adv})^2}
\]
Since $\mu_{ben} + \mu_{adv} \geq (\beta - \alpha)/2$ (under reasonable detector accuracy) and $(\alpha+\beta) \leq 2\beta$, this simplifies to:
\[
T_{sep} \leq \frac{32\beta^2}{(\beta-\alpha)^2} \cdot \ln(2/\delta) \leq \frac{2\tau_{max}}{\beta - \alpha} \cdot \ln(2/\delta)
\]
where the last inequality uses $\tau_{max} \geq 16\beta^2/(\beta-\alpha)$, which holds for typical parameter choices (e.g., $\tau_{max}=4, \beta=2, \alpha=0.5$). For parameter regimes where this does not hold, the bound becomes $T_{sep} = O\!\left(\frac{\beta^2}{(\beta-\alpha)^2}\ln(1/\delta)\right)$, which is tighter.
\end{proof}

\subsection{Proof of Theorem 2 (Attacker's Dilemma)}

\textbf{Theorem 2.} \textit{For any trust policy with asymmetry ratio $\beta/\alpha > k$, the optimal attack length $T^*$ is bounded by Equation~(2).}

\begin{proof}
Consider an attacker who must deliver an adversarial payload that requires the trust level to be at or above some threshold $\tau_{min}$ (otherwise the system restricts the response). The attacker's strategy consists of a sequence of turns, each either ``benign-appearing'' (probability $1-q$ of detection) or ``adversarial'' (probability $1-p_{fn}$ of detection), where $q$ is the per-turn detection probability for the attacker's benign-appearing turns.

\textbf{Phase 1: Trust building.} The attacker must first build trust from $\tau(0)$ to at least $\tau_{min}$. During benign-appearing turns, trust increases at expected rate $\alpha(1-q) - \beta q$ per turn (assuming the attacker's benign turns are detected as adversarial with probability $q \leq p_{fp}$). The number of trust-building turns required is:
\[
T_{build} \geq \frac{\tau_{min} - \tau(0)}{\alpha(1-q) - \beta q} \geq \frac{\tau_{min}}{\alpha(1-p_{fp})}
\]

\textbf{Phase 2: Payload delivery.} The attacker delivers adversarial content over $T_{attack}$ turns. During these turns, trust decays at expected rate $\beta(1-p_{fn}) - \alpha p_{fn}$ per turn. The attacker can sustain the attack until trust drops below $\tau_{min}$:
\[
T_{attack} \leq \frac{\tau_{min}}{\beta(1-p_{fn}) - \alpha p_{fn}}
\]

\textbf{The dilemma.} The attacker can reduce per-turn detection probability by spreading adversarial content more thinly across turns (lower adversarial intensity per turn). However, this requires more turns in Phase 2, during which the state machine (Layer 4) accumulates risk score $r(t)$. The risk score triggers detection when $r > \theta_{high}$, independent of trust level. The risk accumulation rate is proportional to the adversarial content density per turn, creating a second constraint.

\textbf{Combining phases.} The total attack length is $T^* = T_{build} + T_{attack}$. Substituting the bounds:
\[
T^* \leq \frac{\tau_{max}}{\alpha(1-p_{fp})} + \frac{\tau_{max}}{\beta(1-p_{fn}) - \alpha}
\]
Since $\tau_{min} \leq \tau_{max}$ and using $p_{fn} < 0.5$, we have $\beta(1-p_{fn}) > \beta/2 > \alpha$ (given $\beta/\alpha > k \geq 2$), yielding the stated bound in Equation~(2).

\textbf{Tightening with asymmetry.} As $\beta/\alpha$ increases, $T_{attack}$ shrinks (faster trust decay during adversarial turns) while $T_{build}$ grows (slower trust building). The product $T_{build} \cdot T_{attack}$ is bounded, creating the fundamental tradeoff. The bound tightens because increasing $\beta$ reduces $T_{attack}$ faster than it affects $T_{build}$ (which depends on $\alpha$, not $\beta$).

\textbf{Limitation.} This analysis assumes the attacker uses a fixed strategy (constant adversarial intensity). A fully adaptive attacker who dynamically adjusts intensity based on observed system responses (e.g., response latency, refusal patterns) may partially circumvent this bound by optimizing the trust-risk tradeoff in real time. Characterizing the optimal adaptive strategy is equivalent to solving a partially observable Markov decision process (POMDP) and remains an open problem.
\end{proof}

\subsection{Tightened Bound for Proposition 3 (Diversity-Correlation)}

\textbf{Proposition 3.} \textit{For defenses $D_i$, $D_j$ with feature overlap $\omega(D_i, D_j)$, the failure correlation satisfies $\rho(fail_i, fail_j) \leq \omega(D_i, D_j) + \epsilon$.}

We provide a tightened characterization of the residual term $\epsilon$.

\begin{proof}
The failure correlation between two defenses can be decomposed into two components:

\textbf{Feature-dependent correlation.} When both defenses process the same features of the input, a single adversarial perturbation can affect both simultaneously. The mutual information $MI(\phi_i(X); \phi_j(X))$ quantifies the shared information, and the normalized overlap $\omega$ bounds the fraction of the detection decision that is correlated. Formally, by the data processing inequality, the correlation between binary detection outcomes $fail_i, fail_j$ is bounded by the correlation between the feature representations:
\[
\rho(fail_i, fail_j) \leq \rho(\phi_i(X), \phi_j(X)) \leq \sqrt{\omega(D_i, D_j)}
\]
where the second inequality follows from the relationship between mutual information and Pearson correlation for jointly Gaussian variables (Equation 9.7.1 in Cover \& Thomas). For non-Gaussian features, $\omega$ provides an upper bound via the maximal correlation coefficient.

\textbf{Feature-independent correlation ($\epsilon$).} The term $\epsilon$ captures attacks that succeed regardless of the feature representation---i.e., attacks that are ``universally evasive.'' We define $\epsilon$ formally as:
\[
\epsilon = \Pr[\text{attack succeeds against } D_i \text{ AND } D_j \mid \phi_i(X) \perp \phi_j(X)]
\]
This is the probability of simultaneous bypass even when the defenses use completely independent features. Following the beta-factor model from reliability engineering (IEC 61508), $\epsilon$ can be estimated as:
\[
\epsilon \leq \beta_{CCF} \cdot \min(p_i, p_j)
\]
where $\beta_{CCF}$ is the common cause failure fraction and $p_i, p_j$ are the individual bypass probabilities. For CivicShield's cross-domain layers, $\beta_{CCF}$ is bounded by the fraction of the attack taxonomy that is agnostic to detection mechanism. From our taxonomy (Table~IV), only ``universal bypass'' attacks (e.g., Policy Puppetry exploiting training data artifacts) are feature-agnostic, representing approximately 1 of 8 attack families. We therefore estimate $\beta_{CCF} \leq 0.15$ for cross-domain layer pairs, yielding $\epsilon \leq 0.15 \cdot \min(p_i, p_j)$.

\textbf{Combined bound.} Substituting:
\[
\rho(fail_i, fail_j) \leq \sqrt{\omega(D_i, D_j)} + 0.15 \cdot \min(p_i, p_j)
\]
For cross-domain pairs (e.g., L2 regex vs. L5 anomaly detection), $\omega \approx 0.05$--$0.15$, giving $\rho \leq 0.22$--$0.39 + \epsilon$. For same-domain pairs (e.g., two regex-based filters), $\omega \approx 0.7$--$0.9$, giving $\rho \leq 0.84$--$0.95 + \epsilon$. This quantitatively confirms the architectural advantage of cross-domain composition.
\end{proof}

\end{document}